\documentclass[prb,twocolumn,showpacs,amsmath,amssymb,prl]{revtex4-1}
\usepackage{graphicx}  
\usepackage{dcolumn}   
\usepackage{bm}        
\usepackage[utf8]{inputenc}
\usepackage{graphics}
\usepackage{latexsym}
\usepackage{amsmath}
\usepackage{float}
\usepackage{cancel}
\usepackage{pstricks,pst-plot}

\renewcommand{\vec}[1]{\mathbf{#1}}

\newcommand{\ph}{\mathrm{ph}}

\definecolor{mygreen}{rgb}{0.,0.55,0.}

\begin{document}

\title{Brightening odd-parity excitons in transition-metal dichalcogenides: Rashba, Skyrmions, and cavity polaritons}
\author{Luca Chirolli}
\email{l.chirolli@berkeley.edu}
\affiliation{Department of Physics, University of California, Berkeley, CA-94720}
\affiliation{Instituto Nanoscienze-CNR, I-56127 Pisa, Italy}
\date{\today}

\begin{abstract}
Odd-angular momentum exciton states are dark to light in monolayers of transition-metal dichalcogenides and can be addressed only by two-photon probes. Besides, $2p$ excitons states are expected to show a fine splitting that arises from the peculiar electronic band structure of the material, characterized by a finite Berry curvature around each valley.  In this work we study in detail the coupling of photons to $p$ exciton states and we find that the Rashba spin-orbit interaction or a Skyrmion in the transition-metal dichalcogenide substrate can be exploited to engineer finite optical selection rules. The basis mechanism relies on a matching of the exciton angular momentum with the winding of the Rashba spin-orbit interaction or the Skyrmion topological charge. In a photonic cavity, the coupling is enhanced due to photon confinement and the resulting polaritonic branches acquire a mixing with $2p^\pm$ excitons, thus providing a useful tool to detect exciton fine-splitting and and enable novel uses of odd-parity dark excitons. 
\end{abstract}

\maketitle

\section{Introduction}

The generic impact of dimensionality and crystal symmetries in the excitonic spectrum and the associated optical selection rules in two-dimensional (2D) crystals has recently become evident. The rich band structure displaying band topology and winding numbers has been shown to be at the origin of novel selection rules for optical transitions and spectral modifications  in 2D crystals\cite{cao2018unifying,zhang2018optical}. Among different instances of two-dimensional material\cite{roldan2017theory}, monolayers of group VI semiconductor transition-metal dichalcogenides (TMDs) have shown remarkable optical properties, that led to a grown interest in excitons in these materials\cite{ugeda2014giant,MuellerMalic,yu2015valley,manzeli20172d,wang2018colloquium,hsu2019thickness-dependent}. Below the gap the strong Coulomb interaction leads to a dense series of exciton levels\cite{wang2018colloquium}, whose most prominent peaks are the so called A and B excitons. Besides, several other $s$-wave exciton lines have been detected, that strongly deviate from the expected hydrogenic series \cite{he2014tightly,chernikov2014exciton,liu2019magnetophotoluminescence}.  Higher angular momentum exciton states are also predicted to occur\cite{qiu2015nonanalyticity,wu2015exciton}, such as odd-parity $p$-wave states, that are dark to light, due to a mismatch of quantum numbers, and are expected to show a splitting between $2p^+$ and $2p^-$ due to a non-zero Berry curvature effect\cite{srivastava2015signatures,zhou2015berry,wu2015exciton}, although a clear splitting has not been conclusively reported\cite{ye2014probing,wang2015giant}. Nevertheless, being basically optically inactive, $2p$ states are expected to have very few and slow depolarizing channels and can represent an optimal tool for studying valley coherence. These considerations open the way to the engineering of novel selection rules by playing with the fundamental topological properties of the electronic carriers and the possibility to suitably tailor light-matter interaction to generate novel topological polaritons\cite{karzig2015topological,gutierrez2018polariton,latini2019cavity}. 
 
\begin{figure}[b]
\includegraphics[width=0.45\textwidth]{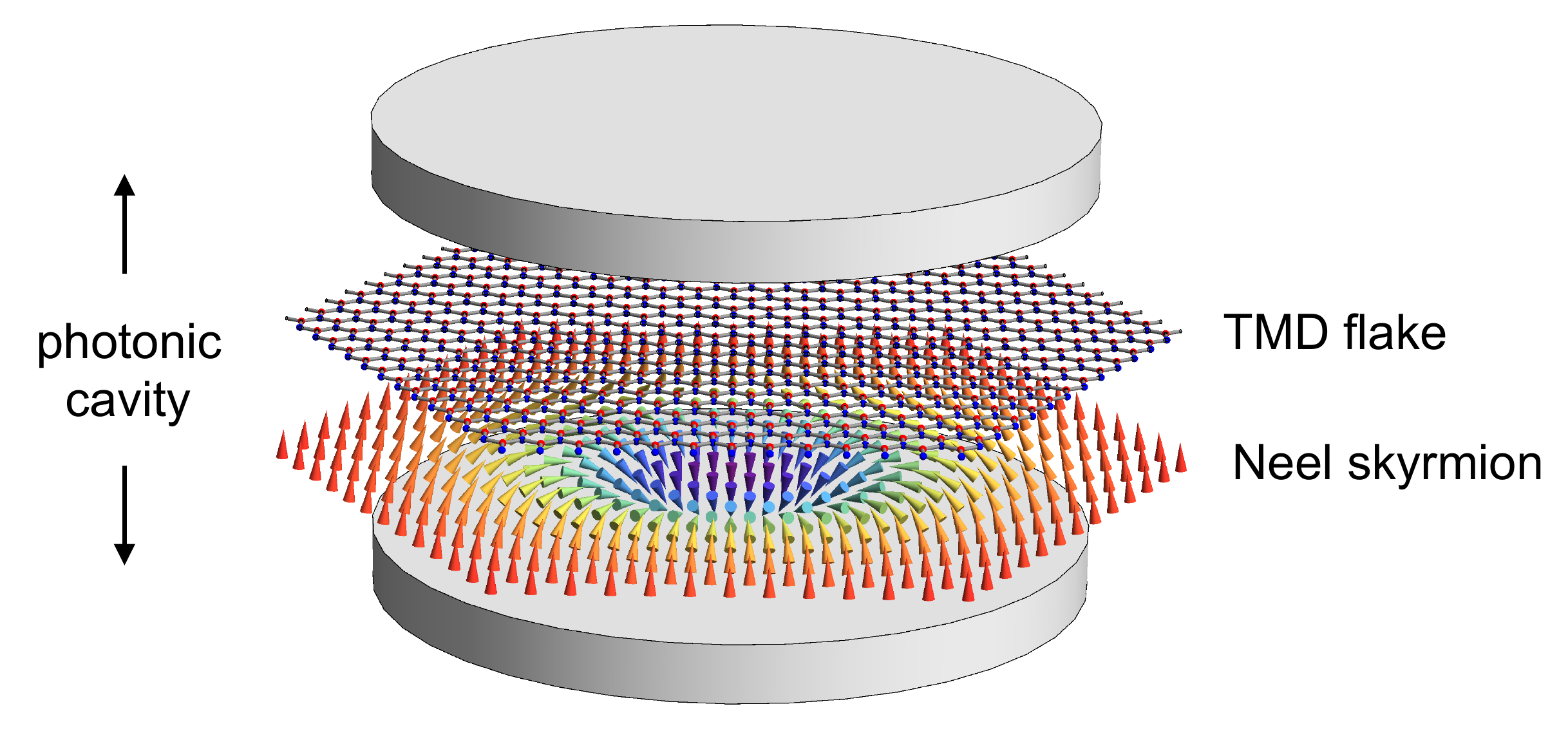}
\caption{Schematics of the device: a flake of TMD is deposited on a substrate containing a Skyrmion and the system is embedded in an optical cavity. 
\label{fig1}}
\end{figure}

The quest for topological states of matter have prompted the relevant role of the Rashba spin-orbit interaction (SOI) as a fundamental ingredient capable to provide the electronic carriers with a topologically non-trivial character (see Refs.~[\onlinecite{hasan2010colloquium,qi2011topological,ando2013topological}] for a review and references therein). The Rashba SOI arises in quasi 2D systems due to the breaking of the in-plane mirror symmetry, $z\to -z$ for systems confined along the $z$ direction. An electric field applied orthogonally to the TMD monolayer can generate a finite Rashba SOI, although the associated spin splitting is very small. Due to the very large intrinsic spin-orbit gap in the valence band, the Rashba SOI in TMDs is appreciable only in the conduction band\cite{kormanyos2014spin-orbit}.

An analogous object capable of breaking spin degeneracy and simultaneously imparting a winding to the electronic carriers is provided by a Skyrmion\cite{nagaosa2013topological,fert2017magnetic}.  Skyrmions are magnetic coplanar spin textures characterized by a topological charge, that arise in magnetic systems by virtue of the competing tendency of neighbouring spins to align perpendicularly or parallel, that favours a chiral order\cite{fert2013skyrmions,fert2017magnetic,zhou2018magnetic}. Several distinct mechanisms can lead to the formation of Skyrmions, such as the Dzyaloshinskii-Moryia interaction or dipolar interaction\cite{nagaosa2013topological,fert2017magnetic}, and the typical size can range from nm to $\mu$m. Based on a mapping between spatially varying in-plane magnetic fields and the Rashba spin-orbit interaction\cite{braunecker2010spin-selective,klinovaja2013spintronics}, a configuration realized by a graphene flake on top of a substrate containing the Skyrmion has been shown to effectively generate a Rashba SOI on the top layer carriers\cite{finocchiaro2017electrical}. Skyrmions have been stabilized both in bulk structures\cite{jonietz2010spin,yu2012skyrmion} and in thin films, interfaces, and multi-layer structues\cite{romming2013writing,moreau-uchaire2016additive,ajejas2018unraveling,ajejas2017tuning}.

In this work we study how to engineer a finite couple between photons and odd-parity $p$ exciton states. We study how to address the predicted $2p^\pm$ exciton splitting in TMDs and look for a way to resolve it experimentally by single-photon probes. We show that the winding of the spin imparted by the Rashba SOI to the carriers, combined with an in-plane magnetic field, allows coupling of light to $2p^\pm$ excitons states, albeit with very small strength due to the weak Rashba SOI. We then suggest to introduce a Skyrmion in the substrate of the system.  The latter generates a spatially varying Zeeman field for the carriers in the TMD monolayer that is rotational invariant along the $z$ direction. Analogously to the case of Rashba SOI, a Skyrmion characterized by a unit winding number can modify the selection rules and mediate the coupling between photons and $p$ excitons through its topological charge, that can match the internal angular momentum of the excitons.

Since the coupling is expected to be very small, we suggest to embed the monolayer in an optical cavity (see Fig.~\ref{fig1}) that allows to enhance the light-matter coupling strength and to study finite momentum deviations from the optical selection rules. The results thus naturally extends to systems with cavity polaritons\cite{pellegrino2014theory,karzig2015topological,gutierrez2018polariton,latini2019cavity}.

The work is structured as follows: in Sec.~\ref{Sec:ExcitonModel} we introduce the excitonic model, by first considering a minimal gapped Dirac Hamiltonian describing low energy physics at the $K/K'$ points in the Brillouin zone in TMDs, and by introducing the exciton wave functions. In Sec.~\ref{Sec:OSR} we introduce the light-matter coupling and discuss optical selection rules for $2p$ excitons including $C_3$ terms in the Hamiltonian. 
In Sec.~\ref{Sec:p-splitting} we address the $2p^\pm$ exciton splitting by first studying the effect of a Rashba SOI and then considering a setup including a Skyrmion in the substrate. We derive the novel selection rules and calculate the relevant matrix elements. In Sec.~\ref{Sec:spin-unlike} we consider the case of spin-unlike exciton and derive the matrix element for coupling of polarized photons to $2p^\pm$ excitons.   In Sec.~\ref{Sec:LMint} we introduce cavity photons and describe the light-matter coupling at finite momentum and in Sec.~\ref{Sec:polaritons} we discuss the resuylting polariton branches and the visibility of the $2p^\pm$ exciton states in the photon spectral function. In Sec.~\ref{Sec:conclusions} we conclude the work with a summary and some final remarks.

\section{Exciton model}
\label{Sec:ExcitonModel}

We consider 2D excitons in a monolayer TMD. The latter has a direct semiconductor gap between conduction and valence bands located at the $K$ and $K'$ points of the Brillouin zone. Excitons in valley $K/K'$ ($\tau=\pm 1$) at finite momentum ${\bf q}$, principal quantum number $n$, and angular momentum $\ell$ are described by the state
\begin{equation}\label{ExcitonPhi}
|\Phi^{n,\ell}_{\tau,{\bf q}}\rangle=\sum_{\bf k}\Phi^{n\ell,\tau}_{{\bf k},{\bf q}}c^\dag_{c,\tau,{{\bf k}+{\bf q}/2}}c_{v,\tau,{\bf k}-{\bf q}/2}|0\rangle,
\end{equation}
where $c_{\alpha,\tau,{\bf k}}$ ($c^\dag_{\alpha,\tau,{\bf k}}$) are fermionic annihilation (creation) operators that remove (add) an electron in the Bloch state $|u^{\alpha,\tau}_{\bf k}\rangle$ in valley $\tau$, band $\alpha=c,v$, and momentum ${\bf k}$, $|0\rangle$ is the neutral filled Fermi sea, and $\Phi^{n\ell,\tau}_{{\bf k},{\bf q}}$ is the ${\bf k}$-space exciton wavefunction. The latter is determined by the electronic Hamiltonian and the Coulomb interaction.  

The minimal model describing a monolayer TMD is given by a gapped Dirac Hamiltonian and accounts for a direct gap located at the $K$ and $K'$ points, as well as direct transitions between conduction and valence band.  It has been recently shown \cite{glazov2017intrinsic,zhang2018optical} that the $C_3$ crystal symmetry allows mixing of even-odd parity selection rules. This can be seen by noticing that the rich band structure of the TMDs allows interband transitions at finite momentum via second order processes involving bands higher in energy. Once projected on the conduction and valence bands, these processes result in trigonal warping terms, a reflection of the $C_3$ symmetry of the crystal  \cite{kormanyos2013monolayer,kormanyos2014spin-orbit,kormanyos2015kp}. We then consider an effective two-band model, that at the $K$ point reads ($\hbar=1$)
\begin{equation}\label{Heff}
H_0({\bf k})=\left(\begin{array}{cc}
\Delta & v_F\left(k_-+\lambda_{\rm tw} k_+^2\right)\\
v_F\left(k_++\lambda^*_{\rm tw}k_-^2\right) & -\Delta
\end{array}\right),
\end{equation}
with $k_\pm=k_x\pm i k_y$. The Hamiltonian at the $K'$ point is obtained by replacing $k_\pm\to -k_\mp$ and $\lambda_{\rm tw}\to \lambda^*_{\rm tw}$. Due to its microscopic origin the scale $\lambda_{\rm tw}$ is on order of the lattice constant\cite{kormanyos2013monolayer}. 

Additionally, TMDs are characterized by a strong intrinsic SOI, that spin-splits the conduction and valence bands and it is at the origin of the A and B exciton peaks in the photoluminescence spectra. It is described by the term
\begin{equation}\label{Hsoi}
H_{\rm SOI}=-\frac{\tau s_z}{4}\left[\delta_c(\openone+\sigma_z)-\delta_v(\openone-\sigma_z)\right],
\end{equation}
with $\delta_c,\delta_v>0$ for the case of MoS$_2$ and MoSe$_2$. The valence band splitting is much larger than the conduction band one. Taking as reference the values determined in Ref.~[\onlinecite{kormanyos2013monolayer}], we have that $\delta_v=146~{\rm meV}$ and $\delta_c=6~{\rm meV}$ for MoS$_2$.  The strong character of the intrinsic SOI promotes the eigenvalue $s$ of the $s_z$ spin component to be a good quantum number and locks spin and valley. Depending on the sign of the conduction band spin splitting, spin-like excitons are the ground states (for Mo-based compounds) or the excited states (for W-based compounds). Focusing on A excitons, that involve the upper most spin-split valence band, allows us to set $\tau s=1$.

The exciton wavefunction $\Phi^{n\ell,\tau}_{{\bf k},{\bf q}}$ entering the state Eq.~(\ref{ExcitonPhi}) is solution of the Bethe-Salpeter equation, that is obtained by projecting the full Hamiltonian including the Coulomb interaction onto two-particle states $c^\dag_{c,{\bf k}+{\bf q}/2}c_{c,{\bf k}-{\bf q}/2}|0\rangle$ (see Appendix~\ref{app:BSE}). Assuming a bare Coulomb interaction $V_q=\frac{2\pi e^2}{\varepsilon q}$, characterized by the material dielectric  constant $\varepsilon$, the exciton eigenstates are given by the 2D hydrogen wavefunctions \cite{yang1991analytic,prada2015effective-mass}, with an effective Bohr radius  $a_{\rm B}\sim 1~{\rm nm}$. 

The 2D hydrogen wavefunctions capture only partially the problem. First of all, the finite thickness of the monolayer results in a screening of the Coulomb interaction, that is described by a momentum dependent dielectric function $\varepsilon(q)=\varepsilon(1+r_0q)$, with $r_0$ a length scale on order of the monolayer thickness\cite{chernikov2014exciton,he2014tightly,wu2015exciton}. This generates matrix elements between excitonic states characterized by different principal quantum number $n$ and a consequent deviation from the $1/(n-1/2)^2$ 2D hydrogen series. 

Secondly, excitons in TMDs have been shown to experience a strong exchange interaction. The latter appears in two ways: i) its short-range contribution shifts the energy of bright excitons with respect to dark spin unlike excitons\cite{echeverry2016splitting}, ii) at finite center of mass momentum ${\bf q}$ it produces a coupling between the $s$ states in different valley, that results in a linear in momentum splitting of the exciton lines at finite momentum\cite{yu2014dirac,wu2015exciton}. This coupling is shown to be zero for odd-angular momentum exciton states (see Appendix~\ref{app:BSE}). 

Furthermore, Berry curvature effects originating from the microscopic Hamiltonian $H_0$ Eq.~(\ref{Heff}), have been shown to be at the origin of fine structure splittings in the exciton spectrum of TMDs \cite{srivastava2015signatures,zhou2015berry}. Due to the pseudo-spinorial character of the eigenstates of the microscopic Hamiltonian $H_0$, a correction to the Coulomb potential arises, that produces a splitting between $2s$ and $2p$ solutions and a further fine splitting between $2p^\pm$ states, both proportional to the Berry curvature of the bands $\Omega_0\equiv\Omega({\bf q}=0)=(v_F/\Delta)^2$. Finally, the presence of the trigonal warping terms gives rise to a mixing between exciton states differing by $\pm 3$ units of angular momentum. This effect is expected to be very small and can be safely neglected.

\section{Optical selection rules}
\label{Sec:OSR}

The Hamiltonian for the light-matter coupling is as usual obtained by minimal coupling substitution ${\bf k}\to {\bf k} +e {\bf A}$ in Eq.~(\ref{Heff}). The vector potential ${\bf A}$ couples at first order to the velocity $V^{\tau}_i=\partial H^{\tau}({\bf k})/\partial {k_i}$ and the Hamiltonian for the light-matter coupling in valley $\tau$ reads
\begin{equation}\label{HamLM}
H_{\rm lm}=\frac{1}{2}\sum_{\tau,i=x,y}\int d{\bf r}\hat{\psi}_\tau^\dag({\bf r})\{A_i({\bf r}),V_i^\tau({\bf r})\}\hat{\psi}_\tau({\bf r})
\end{equation}
with $\hat{\psi}_\tau({\bf r})=\sum_{{\bf k},\alpha}e^{i{\bf k}\cdot{\bf r}}u^{\tau,\alpha}_{\bf k}c_{\tau,\alpha,{\bf k}}$ electronic field operators, with $u_{\bf k}^{\tau,\alpha}$ eigenstates of the Hamiltonian Eq.~(\ref{Heff})$, \{A,B\}=AB+BA$, and   
\begin{eqnarray}
V^\tau_x({\bf k})&=&v_F\left[\tau\sigma_x+2\lambda_{\rm tw}(k_x\sigma_x-\tau k_y\sigma_y)\right],\\
V^\tau_y({\bf k})&=&v_F\left[\sigma_y-2\lambda_{\rm tw} (\tau k_x\sigma_y+k_y\sigma_x)\right].
\end{eqnarray}
The matrix element of the current operator between a photon of polarisation $\nu$ and an exciton with zero momentum ${\bf q}=0$ and quantum numbers $n$ and $\ell$ in valley $\tau$ is given by
\begin{equation}\label{Gamma}
D^{n,\ell}_{\nu,\tau}=ev_F\sum_{\bf k}\Phi^{\tau,n\ell}_{\bf k}\langle u^{v,\tau}_{{\bf k}}|
\hat{P}_{\nu,\tau}|u^{c,\tau}_{{\bf k}}\rangle.
\end{equation}
where $\hat{P}_{\nu,\tau}=\sqrt{2}\left(\tau \sigma_{-\nu\tau}+2\lambda_{\rm tw} k_\nu\sigma_{\nu\tau}\right)$ with $\sigma_{\nu\tau}=(\sigma_x+i\nu\tau \sigma_y)/2$. 

For $\lambda_{\rm tw}=0$ we recover the characteristic optical selection rules, that selectively couple polarized photons to $s$ excitons satisfying $\nu=\tau$ and $d$ excitons satisfying $\nu=-\tau$.   In particular, $D^{n,\ell}_{\nu,\tau}=0$ for $\ell$ odd. 

Photons impinging on the monolayer orthogonally have their electric field in the plane and can couple to excitons with in-pane electric dipole. The resulting optical conductivity reads
\begin{equation}
\sigma=-4\sigma_0\sum_{n,\ell}\left|\bar{D}^{n,\ell}\right|^2E_b{\rm Im}\left[\frac{1}{\omega-\epsilon_{n,\ell}+i\eta}\right],
\end{equation}
where $\sigma_0=e^2/h$, $\epsilon_{n\ell}$ is the energy of the exciton with quantum numbers $n$ and $\ell$, $E_b=h v_F/a_B$ the binding energy, $\eta$ a phenomenological exciton lifetime, and $\bar{D}^{n,\ell}=D^{n,\ell}/D_0$ the adimensional matrix element, with $D_0=ev_F/a_B$ the oscillator strength between cavity photons and $1s$ excitons.

\subsection{Trigonal warping}

The rich band structure of the TMDs allows finite coupling via transitions through bands higher in energy. The additional $k_\pm$ dependence of the velocity operator $\hat{P}_{\nu,\tau}$ at finite $\lambda_{\rm tw}$ can compensate the angular momentum of $2p$ excitons and the matrix elements $D^{n,\ell}_{\nu,\tau}$ acquires finite value for $\ell=\pm 1$. The linear dependence in $\lambda_{\rm tw}k$ produces, once integrated over the exciton wave function, a coefficient $\lambda_{\rm tw}/a_B$. Second order corrections, that can be obtained by expanding both conduction and valence band eigenstates at first order $\lambda_{\rm tw} \xi k^2$, with $\xi=v_F/\Delta$, necessarily bring a factor $(\lambda_{\rm tw}\xi/a_B^2)^2$, and are suppressed.

\begin{equation}\label{p-selection-rules}
D^{2,\ell}_{\nu,\tau}=D_{\rm tw}\delta_{\nu,-\tau}\delta_{\nu,-\ell},\qquad D_{\rm tw}=D_0\frac{\lambda_{\rm tw}}{a_B}c_{2p},
\end{equation}
where $c_{2p}=8/(9\sqrt{6})$ is a form factor associated to the ratio between $2p$ and $1s$ exciton Bohr radia. Using the parameters of Ref.~[\onlinecite{wu2015exciton}], with an exciton Bohr radius in MoS$_2$ on order of $a_B\sim 0.277~{\rm nm}$, realistically gives $\lambda_{\rm tw}/a_B \simeq 0.1$. On the other hand, for an effective exciton radius of $a_{\rm B}\simeq 1~{\rm nm}$, the coupling is reduced. 

This result marks a breaking of the optical selection rules, as anticipated in Refs.~[\onlinecite{glazov2017intrinsic,zhang2018optical}], that owes to the $C_3$ symmetry of the crystal Bloch functions at the $K$ and $K'$ points. Finite coupling to $p$ excitons can be  understood by noticing that, upon neglecting the linear in ${\bf k}$ term, the Hamiltonian Eq.~(\ref{Heff}) is equivalent to gapped bilayer graphene, for which optical selection rules predict bright $p$ excitons and dark $s$ excitons. Thus, the $C_3$ crystal symmetry yields a mixing of even-odd parity selection rules.

\section{Resolving the $2p^\pm$ splitting}
\label{Sec:p-splitting}

If on the one hand the selection rule Eq.~(\ref{p-selection-rules}) allows coupling to exciton $2p$-states, it does not allow to resolve the $p^\pm$ splitting predicted by finite Berry curvature effects. This is due to the fact that out of the four $2p$ states, $2p^\pm$ in valley $K$ and $2p^\mp$ in valley $K'$, that are related by time-reversal symmetry and are pair-wise degenerate, only two of them ($2p^+$ in valley $K$ and $2p^-$ in valley $K'$) become bright according to Eq.~(\ref{p-selection-rules}). We now show that the inclusion of a Rashba SOI or the presence of a Skyrmion on the substrate allow to couple the four exciton $2p$ states to the light field. 

Before proceeding we adapt the notation to account for the spin degree of freedom. The spin $s$ eigenstates of the electronic Hamiltonian $H({\bf k})=H_0({\bf k})+H_{\rm SOI}$ are denoted as $|u^{\alpha,\tau}_{{\bf k},s}\rangle$. The presence of the strong SOI locks spin and valley, so that at valley $\tau$ the lowest energy transition is between conduction and valence band states of spin $s=\tau$, so that selection rules for coupling to light of spin-like excitons is generally written in terms of the matrix element $\langle u^{v,\tau}_{{\bf k},\tau}|\hat{P}_{\nu,\tau}|u^{c,\tau}_{{\bf k},\tau}\rangle$. We now consider the modification of $D^{n,\ell}_{\nu,\tau}$ due to the presence of Rashba SOI and the presence of a Skyrmion.

\subsection{Rashba spin-orbit interaction}
\label{Sec:Rashba}

Let us consider a Rashba SOI, that can be obtained by sandwiching the monolayer TMD between different materials, that generate a breaking of the $z\to -z$ symmetry, and can be enhanced by application of an out-of-plane electric field. We further apply an in-plane magnetic field that couples to the electronic degrees of freedom via the Zeeman interaction, so that we add the following term $H_1=H_R+H_Z$, 
\begin{equation}\label{Eq:Rashba}
H_1=\lambda_R(\tau s_x\sigma_y-s_y\sigma_x)+\frac{1}{2}g\mu_BBs_x,
\end{equation}
with $g$ the TMD $g$-factor and $\mu_B$ the Bohr magneton. 
We consider the Rashba and Zeeman terms as perturbations and calculate the electronic eigenstates at first order perturbation theory in $H_1$. This way, for instance in valley K ($\tau=1$) we have
\begin{equation}\label{Psi1stH1}
|\psi^{\alpha,K}_{{\bf k},\uparrow}\rangle^{(1)}=|\psi^{\alpha,K}_{{\bf k},\uparrow}\rangle +\frac{1}{\delta_\alpha}|\psi^{\alpha,K}_{{\bf k},\downarrow}\rangle\langle\psi^{\alpha,K}_{{\bf k},\downarrow}|H_1|\psi^{\alpha,K}_{{\bf k},\uparrow}\rangle.
\end{equation}
The matrix elements of $H_1$ between the unperturbed states are given, in a unified the notation, by
\begin{eqnarray}
\langle\psi^{c,\tau}_{{\bf k},-\tau}|H_1|\psi^{c,\tau}_{{\bf k},\tau}\rangle&=&-i\lambda_RS_ke^{i\tau\phi_k}+V_Z,\\
\langle\psi^{v,\tau}_{{\bf k},-\tau}|H_1|\psi^{v,\tau}_{{\bf k},\tau}\rangle&=&i\lambda_RS_ke^{i\tau\phi_k}+V_Z,
\end{eqnarray}
with $S_k=v_F k/\sqrt{(v_Fk)^2+\Delta^2}$, $V_Z=g\mu_B B/2$, where a crucial winding $e^{i\tau\phi_k}$ appears due to the Rashba field. 

\begin{figure}[t]
\includegraphics[width=0.45\textwidth]{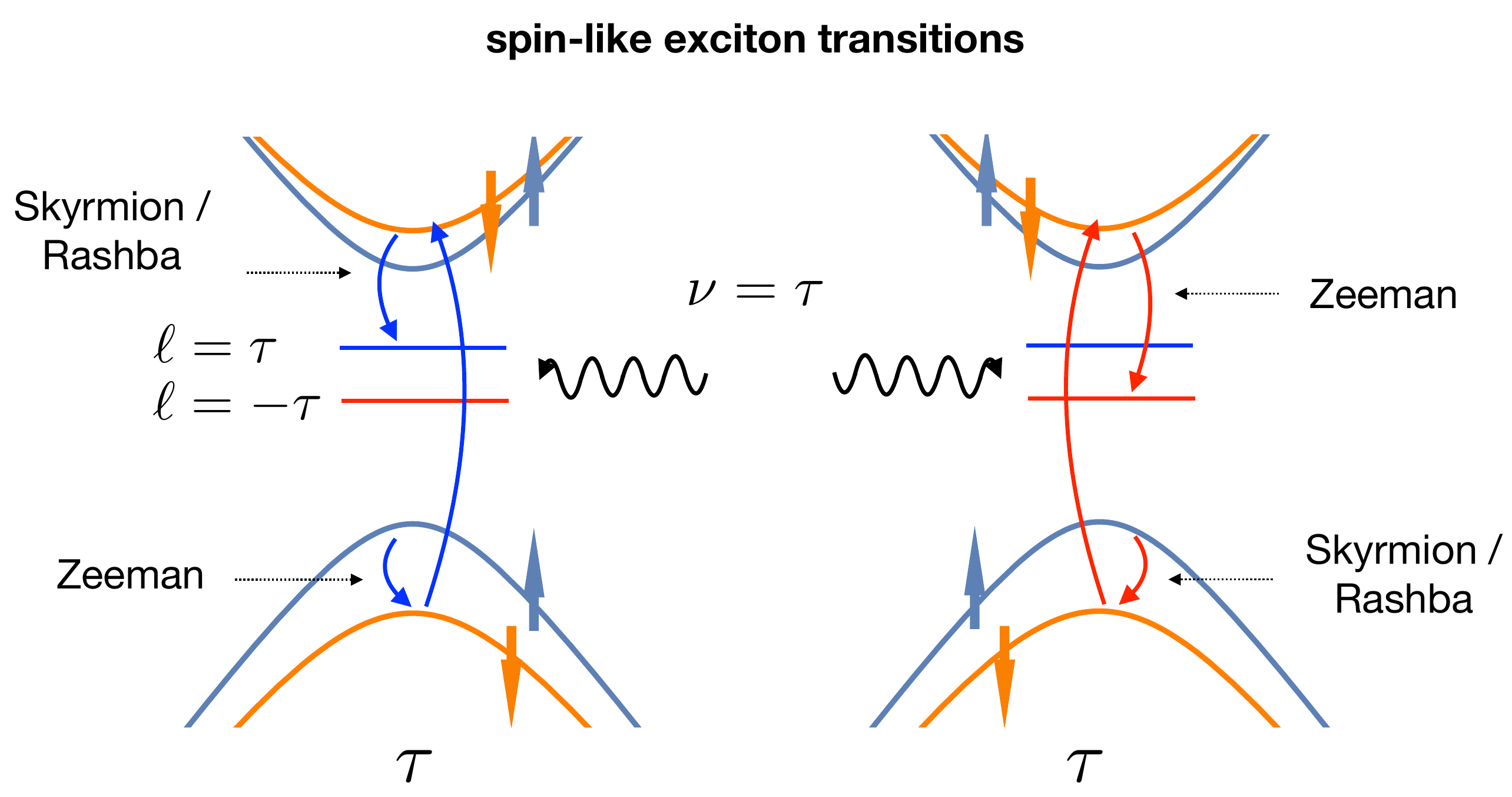}
\caption{Schematics of the coupling between polarization $\nu$ photons and spin-like $p$ excitons in valley $\tau$. A $N=1$ Skyrmion or the Rashba field can simultaneously mediate a spin-flip transition and an angular momentum quantum necessary to couple to $p$ excitons. The left and right schemes refer to the two possible transition to the two $2p^\pm$ states, in which the Zeeman and Rashba/Skyrmion term exchange their role. 
\label{fig2}}
\end{figure}

It then follows that the matrix element between conduction and valence band $\langle u^{v,\tau}_{{\bf k},\tau}|\hat{P}_{\nu,\tau}|u^{c,\tau}_{{\bf k},\tau}\rangle$ can be expanded at second order in the perturbation $H_1$, by expanding the bra at first order in the Zeeman field and the ket at first order in the Rashba field, and viceversa. This way, a non-zero matrix element results between conduction and valence band, as schematically depicted in Fig.~\ref{fig2}, that is at first order in the Zeeman field and in the Rashba field, the latter carrying a finite winding that can compensate the exciton angular momentum of the $p$ states. The resulting new selection rules read
\begin{equation}\label{GammaRashba}
D^{2,\ell}_{\nu,\tau}=-i\tau D_R(\delta_{\tau,\ell}+\delta_{\tau,-\ell}),
\end{equation}
where the coupling strength is 
\begin{equation}\label{GammaR}
D_R=D_0\frac{\lambda_RV_Z}{\delta_c\delta_v}\frac{\hbar v_F}{a_B\Delta}\frac{4\sqrt{2}}{9\sqrt{3}}.
\end{equation}
The two terms appearing in Eq.~(\ref{GammaRashba}) proportional to $\delta_{\tau,\pm \ell}$ are obtained by processes in which the conduction band eigenstate Eq.~(\ref{Psi1stH1}) is expanded at first order in the Rashba field and the valence band at first order in the Zeeman term, and viceversa, as schematically depicted in Fig.~\ref{fig2}. 

\begin{figure}[t]
\includegraphics[width=0.45\textwidth]{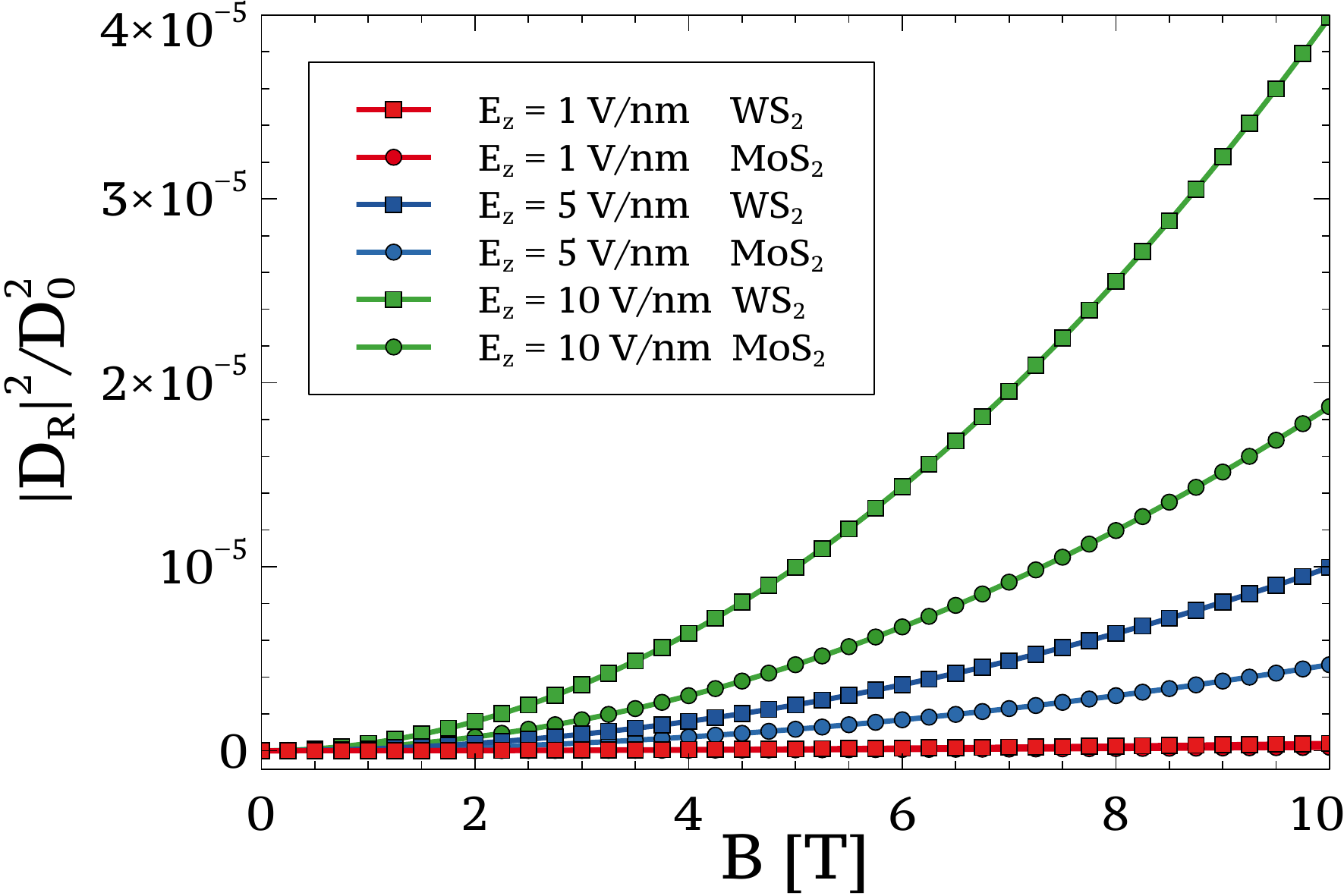}
\caption{Square modulus of the matrix element $D_R$ due to the Rashba SOI, in units of $D_0^2$, for different value of applied electric field $E_z$, for two TMDs, MoS$_2$ (circle markers) and WS$_2$ (square markers), characterized by electric dipoles $d_z=0.02$ $e$\AA ~and $d_z=0.06$ $e$\AA, respectively.
\label{figR}}
\end{figure}

Assuming that an electric field is applied orthogonal to the monolayer, in Fig.~\ref{figR} we plot the square modulus of the Rashba-induce matrix element $D_R$, that represents the magnitude of the signal in the optical conductivity. Curves are shown for MoS$_2$ and WS$_2$, for different values of the electric field $E_z$. The latter is related to the Rashba coupling constant as $\lambda_R=E_zd_z$, and values of the electric dipole has been taken from Ref.~[\onlinecite{slobodeniuk2016spin-flip}]. Clearly, a very strong electric field is required to overcome the transition through the valence band, that involves the spin-splitting $\delta_v$, and have an appreciable coupling. 

\subsection{Skyrmions}

In order to overcome the difficulty posed by the requirement of a strong Rashba coupling constant, we consider to employ spatially varying magnetic fields\cite{braunecker2010spin-selective,klinovaja2013spintronics}. It has been estimated that a Rashba coupling constant of up to 10 meV can be realized for an array of equally spaced nanomagnets \cite{karmakar2011controlled,kjaergaard2012majorana}  with a periodicity of 100 nm. These numbers are exactly what we need in order to obtain a strong effective Rashba SOI and at the same time separate the length scales of the exction motion and the electronic motion. Unfortunately, the resulting Rashba misses an important in-plane spin component and the array of spatially periodic nanomagnet breaks the rotation symmetry of the system, that is required to effectively couple $p$-excitons to the valence-conduction band transition. We then resort to the use of a Skyrmion placed in the substrate, that has been shown to effectively generate a Rashba SOI on the carriers of a graphene flake placed on top of a Skyrmion\cite{finocchiaro2017electrical}.

The magnetization field of a Skyrmion is described by a classical unit vector ${\bf n}({\bf r})$ that represents a maps from the one-point compactification of the plane (${\mathbb R}^2\cup\{\infty\}\simeq S^2$) onto the two-sphere spanned by ${\bf n}$.  The map ${\bf n}: S^2\to S^2$ is classified by the homotopy class $\pi_2(S^2)\in {\mathbb Z}$ that defines the integer-valued $C{\mathbb P}^1$ topological invariant
\begin{equation}
{\cal Q}\equiv\frac{1}{4\pi}\int dxdy~ {\bf n}\cdot \partial_x{\bf n}\times \partial_y{\bf n}.
\end{equation}
The index ${\cal Q}$ counts the number of times the Skyrmion magnetization ${\bf n}$ wraps around the sphere as it evolves from the Skyrmion centre to infinity. Parametrizing the plane in polar coordinates ${\bf r}=(r,\phi)$ and the unit sphere by $(\Theta,\Phi)$, the unit vector ${\bf n}$ is written as 
\begin{equation}
{\bf n}({\bf r})=(\cos\Phi(\phi)\sin\Theta(r),\sin\Phi(\phi)\sin\Theta(r),\cos\Theta(r)),
\end{equation}
and the mapping is specified by the function $\Phi(\phi)$ and $\Theta(r)$, as long as the magnetization field becomes $\phi$-independent far from the Skyrmion center [\onlinecite{nagaosa2013topological}]. Choosing 
\begin{equation}
\Phi(\phi)=N\phi +\gamma,
\end{equation} 
and $\Theta(r)$ a function ranging from $\pi$ at $r=0$ to 0 as $r\to\infty$ we have that ${\cal Q}=N$. 

The Skyrmion in the substrate generates a spatially varying Zeeman field described by the term $H_S=J{\bf n}({\bf r})\cdot{\bf s}$. Assuming the Skyrmion to vary on a spatial scale much longer than the inverse intervalley separation, it is a good approximation to neglect inter-valley scattering generated by the Skyrmion.  This way, the electronic eigenstates at first order perturbation theory read
\begin{equation}\label{1stPT-Skyrmion}
|u^{\alpha,\tau}_{{\bf k},\tau}\rangle^{(1)}=|u^{\alpha,\tau}_{{\bf k},\tau}\rangle+\frac{1}{\delta_\alpha}\sum_{{\bf k}'}\langle u^{\alpha,\tau}_{{\bf k}',-\tau}|H_S|u^{\alpha,\tau}_{{\bf k},\tau}\rangle |u^{\alpha,\tau}_{{\bf k}',-\tau}\rangle,
\end{equation}
where the spatially varying spin texture couples different momentum states via the matrix element
\begin{eqnarray}
\langle u^{\alpha,\tau}_{{\bf k}',-\tau}|H_S|u^{\alpha,\tau}_{{\bf k},\tau}\rangle&=&J\langle u^{\alpha,\tau}_{{\bf k}'}|u^{\alpha,\tau}_{\bf k}\rangle\nonumber\\
&\times& \int d{\bf r} e^{i({\bf k}-{\bf k}')\cdot{\bf r}}e^{iN\tau\phi}\sin\Theta(r), 
\end{eqnarray}
and the second term in the r.h.s. comes from the matrix element of opposite spin states mediated by the Skyrmion and contains the winding number $N$ multiplied by the spin state quantum number $s=\tau$. 

By further applying also an in-plane magnetic field in the, say, $x$-direction, the matrix element between conduction and valence band $\langle u^{v,\tau}_{{\bf k},\tau}|\hat{P}_{\nu,\tau}|u^{c,\tau}_{{\bf k},\tau}\rangle$ can be expanded at second order in the perturbation, analogously to Sec.~\ref{Sec:Rashba}, by expanding the bra at first order in the Zeeman field and the ket at first order in the Skyrmion field, and viceversa. 

\begin{figure}[t]
\includegraphics[width=0.45\textwidth]{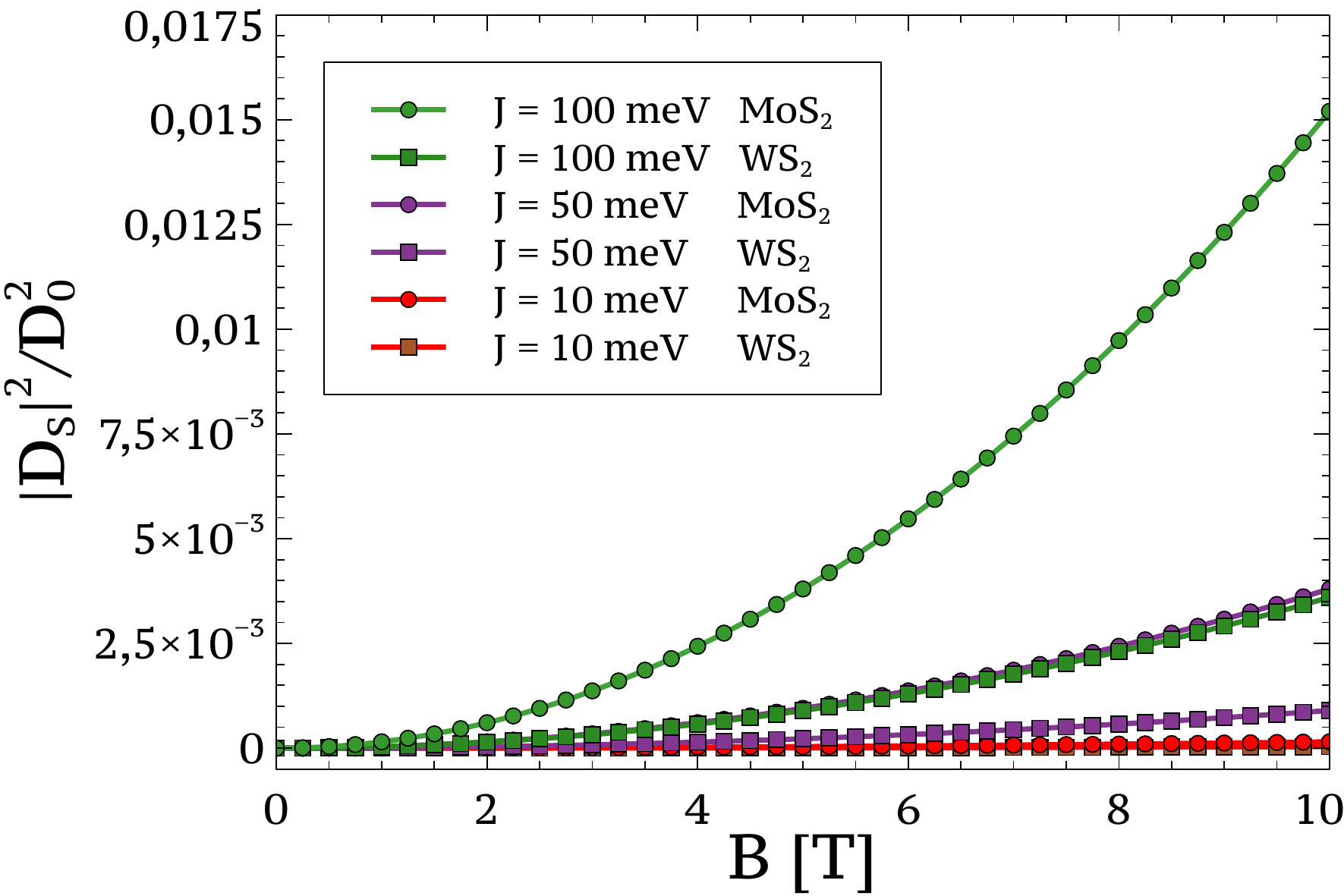}
\caption{Square modulus of the matrix element $D_S$ due to a Skyrmion, in units of $D_0^2$, for different value of the Skyrmion exchange $J$, for two TMDs, MoS$_2$ (circle markers) and WS$_2$ (square markers).
\label{figS}}
\end{figure}

The resulting selection rules are given by
\begin{equation}\label{GammaSkyrmion}
D^{2,\ell}_{\nu,\tau}=i\tau D_S(\delta_{N\tau,\ell}-\delta_{N\tau,-\ell}),
\end{equation}
where the role of the winding number $N$ becomes manifest. This way, for a unit winding $N=1$ we can couple to both $2p^+$ and $2p^-$ state in each valley. The strength of the coupling is 
\begin{equation}
D_S=D_0\frac{JV_Z}{\delta_c\delta_v}c_n,
\end{equation}
that is completely analog to Eq.~(\ref{GammaR}) obtained for the case of Rashba SOI (see Fig.~\ref{fig2}), with the very difference that the exchange $J$ can range from from few meV to 100 meV. The coefficient $c_n\lesssim 1$ depends on the precise form of the function $\Theta(r)$. The square of the matrix element $D_S$ is plotted in Fig.~\ref{figS} versus the applied in-plane magnetic field, for different values of the exchange interaction $J$. The curves are qualitatively similar to those shown in Fig.~\ref{figR}, with the major difference being the scale of the signal, that can be 100 times stronger for an exchange $J\sim 10$ meV.

\section{Spin-unlike Excitons}
\label{Sec:spin-unlike}

So far, along the entire analysis we assumed to deal with spin-like excitons, that form between electronic states with same spin. These are the relevant lowest energy excitons in MoS$_2$ and MoSe$_2$ compounds, where the sign of the splitting in the intrinsic spin-orbit interaction Eq.~(\ref{Hsoi}) is positive. In the compounds formed with the transition metal W, that are WS$_2$ and WSe$_2$, we have $\delta_c<0$. In addition, a strong exchange interaction further shifts up in energy spin like excitons\cite{echeverry2016splitting} (the exchange interaction is zero for spin unlike transitions), so that the lowest energy valence-conduction band electronic transition is between opposite spin states. The spin-orbit interaction in WS$_2$ and WSe$_2$ is  stronger than in Mo-based compounds, with typical values of $\delta_v=180~{\rm meV}$ and $|\delta_c|=10~{\rm meV}$, so that $s_z$ spin component is a good quantum number and the spin-valley locking is guaranteed.  

1$s$ excitons forming between spin unlike states have their dipole matrix element pointing about the out-of-plane direction and do not couple to light perpendicular to the sample. For this reason they are termed dark. They can nonetheless couple to light propagating in-plane with the electric field orthogonal to the plane\cite{robert2017fine,wang2017in-plane,molas2017brightening}. A residual coupling between spin unlike excitons in different valleys occurs via a residual short-range dipole-dipole interaction\cite{slobodeniuk2016spin-flip}, that splits the excitons in inter-valley symmetric and anti-symmetric combinations. This has been rigorously shown via group theory argument to be the case for $s$-wave excitons\cite{dery2015polarization,slobodeniuk2016spin-flip,robert2017fine} and a splitting of $\delta_{\rm ex}\sim 0.6$ meV has been measured experimentally \cite{robert2017fine,molas2019probing}. A natural question occurs, whether or not the same mechanism applies to $p$ excitons. Being the intervalley coupling proportional to the exciton wavefunction at the origin we expect no coupling for $p$ excitons and higher principal quantum number states.

Due to the fact that dark excitons have opposite spin locked to the out-of-plane direction, an in-plane magnetic field has  been proved effective in coupling $s$ excitons to light\cite{molas2017brightening,robert2017fine,molas2019probing}. Analogously, it is clear that a Skyrmion, that contains both an in-plane magnetic field and a topological charge that results in a winding of the Zeeman field, can couple to $p$ excitons at first order perturbation theory.

The calculation goes along the line of the one previously explained for spin-like excitons. Denoting as $|u^{c,\tau}_{{\bf k},s}\rangle$ the bare electronic band eigenstates, the coupling of photons with polarization $\nu$ to spin-unlike excitons with principal quantum number $n$ and angular momentum $\ell$ is given by
\begin{equation}\label{dark}
D_{\rm Unlike}=ev_F\sum_{\bf k}\Phi^{\tau,n\ell}_{\bf k}\langle u^{v,\tau}_{{\bf k},-\tau}|
\hat{P}_{\nu,\tau}|u^{c,\tau}_{{\bf k},\tau}\rangle.
\end{equation}
The matrix element between valence and conduction band is calculated by expanding the band eigenstates at first order perturbation theory in the Skyrmion field as in Eq.~(\ref{1stPT-Skyrmion}) and the result is
\begin{equation}\label{GammaUnlike-S}
D^S_{\rm Unlike}=-i\tau\delta_{\nu,\tau}\left(\delta_{\ell,\tau N}\frac{J}{\delta_c}-\delta_{\ell,-\tau N}\frac{J}{\delta_v}\right)D_0c_{2p}.
\end{equation}
The transition to opposite angular momentum $\ell$ states goes via different bands, resulting in a different oscillator strength. 

For a Skyrmion sufficiently close to the TMD sample, the amplitude $\Gamma_{\rm Unlike}$ is a fraction of the amplitude for bright A excitons $D_0$. In general, in a system containing several Skyrmions, all with the same winding $N$, the signal will be multiplied by the number of Skyrmions (every transition is independent and among them incoherent). 

For the case of Rashba SOI the result can be immediately written by analogy with Eqs.~(\ref{GammaRashba},\ref{GammaSkyrmion},\ref{GammaUnlike-S})
\begin{equation}\label{GammaUnlike-R}
D^R_{\rm Unlike}=-i\tau\delta_{\nu,\tau}\left(\delta_{\ell,\tau}\frac{\lambda_R}{\delta_c}+\delta_{\ell,-\tau}\frac{\lambda_R}{\delta_v}\right)D_0c_{2p},
\end{equation}
where attention has to be paid at the sign of $\delta_c$. It follows that through the application solely of an electric field orthogonal to the system it is possible to optically activate one of the $2p^\pm$ states of spin unlike excitons. Importantly, in systems like WS$_2$ and WSe$_2$ spin unlike excitons represent the lowest energy excitons and without the application of an in-plane magnetic field one can rule out the $s$ excitons, so that the signal comes only from those states.

\subsection{Optical conductivity}

At the level of the optical conductivity the spin-like matrix elements for $p$ exciton activated by Rashba SOI or a Skyrmion are probably too small to be experimentally resolved. On the other hand, for the case of spin-unlike excitons, the first order perturbation theory result Eqs.~(\ref{GammaUnlike-S},\ref{GammaUnlike-R}) can be resolved in the optical conductivity. Furthermore, for the case of an applied electric field orthogonal to the monolayer the Rashba induced matrix element Eq.~(\ref{GammaUnlike-R}) can be switched on and off and the experimental apparatus is much simpler than the case with Skyrmions in the substrate.  

\begin{figure}[t]
\includegraphics[width=0.45\textwidth]{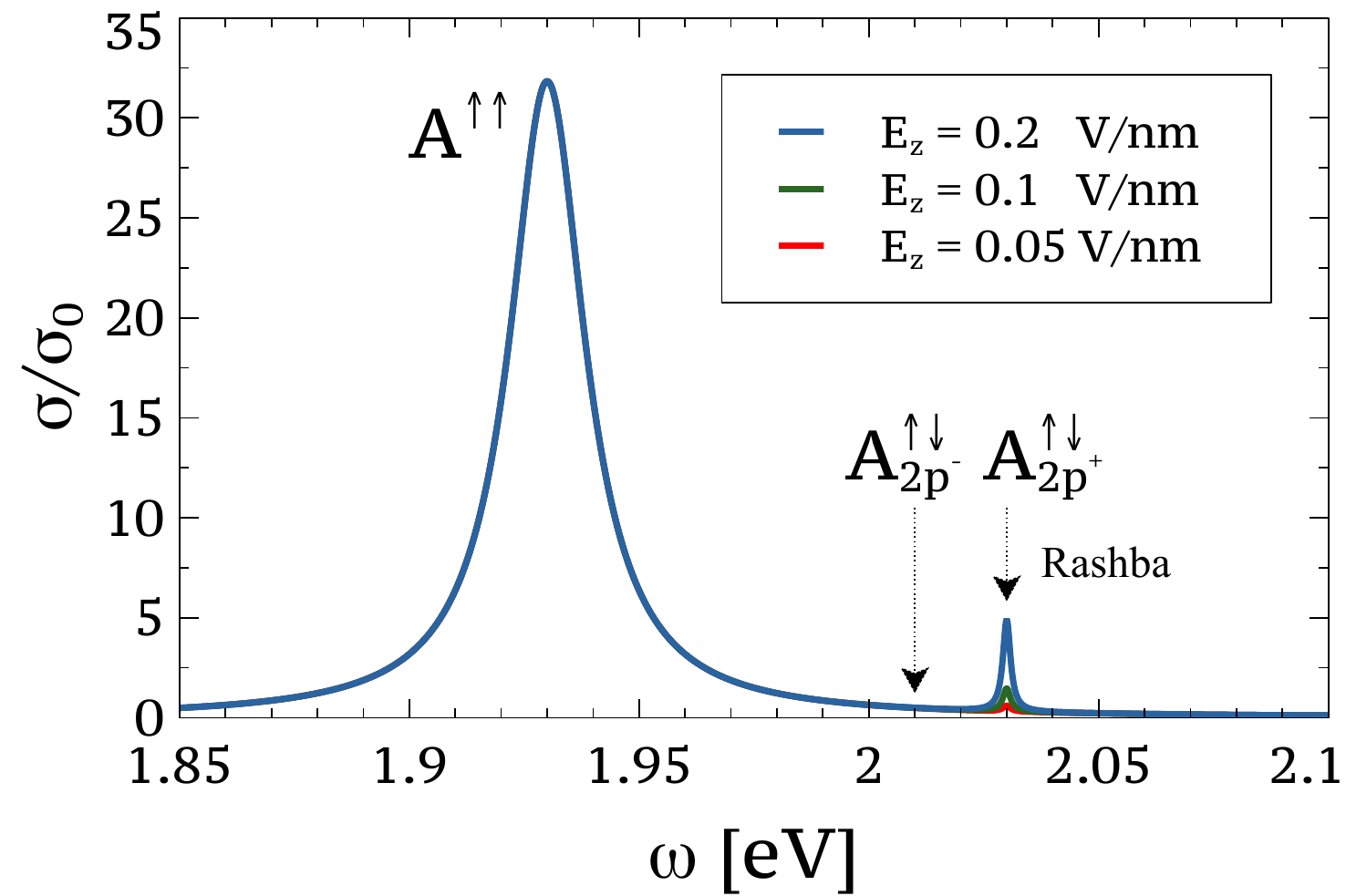}
\caption{Optical conductivity in presence of an orthogonal electric field $E_z$. The position of the A exciton is chosen to be at $\epsilon_A=1.93$ eV and the position of the $2p^\pm$ spin unlike excitons are chose to be at $\epsilon_{2p^-}=2.01$ eV and $\epsilon_{2p^+}=2.03$ eV. The resonances are referred to a given valley $\nu=\tau=1$. Phenomenological exciton lifetimes have been chosen $\eta_A=0.01$ eV and $\eta_{\rm unlike}=0.001$ eV. 
\label{figOC}}
\end{figure}

In Fig.~\ref{figOC} we plot the resulting optical conductivity for several values of the applied electric field, $E_z=0.05, 0.1, 0.2$ V/nm.   Different phenomenological lifetimes have been chosen for spin-like and spin-unlike excitons, that are expected to be much longed lived, $\eta_A=0.01$ eV and $\eta_{\rm unlike}=0.001$ eV. Remarkably, the position of one of the two $2p$ states can be very well experimentally resolved. 

Concerning the other $2p^\mp$ state, that remains dark for $\lambda_R/\delta_v\ll 1$, by comparing the selection rule due to trigonal warping Eq.~(\ref{p-selection-rules}) and the one due to Rashba SOI  Eq.~(\ref{GammaUnlike-R}), we notice that they both promote $\ell=\tau$. Whereas the chirality of the trigonal warping term is dictated by symmetry, the selection rules Eq.~(\ref{GammaUnlike-R}) depends solely on the sign of $\delta_c$. In fact, for $\delta_c>0$ one has to replace $\ell\to -\ell$ in Eq.~(\ref{GammaUnlike-R}). From experiments and theoretical calculations it certainly follows that $\delta_c$ and $\delta_v$ have opposite sign in W-based compounds. However, probably the overall sign can be questioned. If $\delta_v<0$ and $\delta_c>0$ then one can access both the $2p$ states, one via the Rashba SOI and the other via the trigonal warping corrections, provided an in-plane magnetic field is applied to flip the spin. One would then obtain the mixed selection rules
\begin{equation}\label{GammaUnlike-mix}
D^R_{\rm Unlike}=\left(-i\tau\delta_{\nu,\tau}\delta_{-\ell,\tau}\frac{\lambda_R}{\delta_c}+\delta_{\nu,-\tau}\delta_{\tau,\ell}\frac{\lambda_{\rm tw}V_Z}{a_B\delta_c}\right)D_0c_{2p}.
\end{equation}
However, to fully exploit the trigonal warping selection rules one has to overcome the $\delta_c=10$ meV, that requires an in-plane field on order of 100 T.

\section{Cavity Light-matter coupling}
\label{Sec:LMint}

We now consider a planar cavity of height $L_z$ with perfectly conducting mirrors, in which photons have quantised out-of-plane momentum $k_z$ and carry in-plane momentum $\vec{q}=(q_x,q_y)$. In the lowest subband the momentum takes the value $k_z = \pi/L_z$. For a cavity with perfectly conducting mirrors, the TE and TM modes are degenerate. In the basis of circularly polarised light, for which we define the bosonic operators $a_{\nu,\vec{q}}$ with polarisations $\nu = \pm$, the photonic Hamiltonian reads \cite{gutierrez2018polariton}
\begin{align}
    H_{\ph} & =
    \sum_{\nu,{\bf q}}
    \hbar \omega_{\vec{q}}
    a_{\nu,\vec{q}}^\dagger
    a_{\nu,\vec{q}}
    \, ,
    \label{eq:h_ph}
\end{align}
where $\omega_\vec{q}$ denotes the photon frequency. 

Integrating over the spatial distribution of the cavity vector potential, the light-matter coupling Hamiltonian Eq.~(\ref{HamLM}) takes the form
\begin{eqnarray}
H_{\rm lm}&=&iev_F\sum_{\nu,{\bf q}}\sum_{{\bf k},\alpha,\beta}\left|{\cal A}_q\right|\langle u^{\tau,\beta}_{{\bf k}-{\bf q}/2}|
\hat{P}_{\nu,\tau}({\bf q},{\bf k})|u^{\tau,\alpha}_{{\bf k}+{\bf q}/2}\rangle
\nonumber\\
&\times&c^\dag_{\alpha,{\bf k}-{\bf q}/2} c_{\beta,{\bf k}+{\bf q}/2}(a^\dag_{\nu,{\bf q}}-a_{-\nu,-{\bf q}}),
\end{eqnarray}
where $|{\cal A}_q|=1/\sqrt{\hbar\epsilon_0 \mathcal{V}\omega_{\vec{q}}}$ and ${\cal V}$ is the cavity volume. At finite ${\bf q}$, the presence of the cavity forces the TM mode to develop a finite angle with the out-of-plane direction, that yields a deviation from the optical selection rules analogously to the results of Ref.~[\onlinecite{gutierrez2018polariton}], and  we have 
\begin{eqnarray}\label{Eq:Op-LM-q}
\hat{P}_{\nu,\tau}({\bf q})&=&\sqrt{2}\left(\tau \sigma_{-\nu\tau}+2\lambda_{\rm tw} k_\nu\sigma_{\nu\tau}\right)c_q\nonumber\\
&-&\sqrt{2}\left(\tau \sigma_{\nu\tau}+2\lambda_{\rm tw} k_{-\nu}\sigma_{-\nu\tau}\right)s_qe^{-2i\nu\theta_q}.
\end{eqnarray}
where $\tan \theta_q=q_y/q_x$, $c_q=\cos^2(f_q/2)$,  $s_q=\sin^2(f_q/2)$, and $\cos(f_q)=k_z/\sqrt{k_z^2+q^2}$ is the angle between the TM mode and the out-of-plane direction. 

It is clear that the coupling between cavity photons and excitons will be modified from the optical selection rules by the ${\bf q}$ dependence of the light-matter operator Eq.~(\ref{Eq:Op-LM-q}), the exciton wavefunction $\Phi_{{\bf k},{\bf q}}$, and the conduction and valence band eigenfunctions $|u^{\tau,n}_{\bf k}\rangle$.  
The generalization at finite momentum ${\bf q}$ of the light-matter matrix element Eq.~(\ref{Gamma}) reads
\begin{equation}
\Gamma^{n\ell}_{\nu,\tau}({\bf q})=-ev_F{\cal A}_{\bf q}\sum_{\bf k}\Phi^{n\ell,\tau}_{{\bf k},{\bf q}}\langle u^{\tau,v}_{{\bf k}-{\bf q}/2}|
\hat{P}_{\nu,\tau}({\bf q})|u^{\tau,c}_{{\bf k}+{\bf q}/2}\rangle,
\end{equation}
where now we are interested in the Hamiltonian coupling between a photon of polarization $\nu$ and momentum ${\bf q}$ and an exciton with same momentum and quantum numbers $n$ and $\ell$.
 
The dependence on ${\bf q}$ coming from $|u^{\tau,n}_{\bf k}\rangle$ can be shown to be completely negligible, whereas the dependence coming from the light-matter operator $\hat{P}_{\nu,\tau}({\bf q})$ enters as a multiplicative factor to the optical selection rules obtained in the previous sections. This way the exciton photon coupling arising from the trigonal warping terms, the Rashba spin-orbit interaction, and the Skyrmion are simply given by
\begin{eqnarray}
\Gamma^{\rm cav}_{\rm tw}&=&\Gamma_{\rm tw}\left[c_p\delta_{\nu,\bar\tau}\delta_{\bar\nu,\ell}-s_q e^{-i\ell\theta_q}\delta_{\nu,\tau}\delta_{\nu,\ell}\right],\\ \label{p-selection-rules-cav}
\Gamma^{\rm cav}_R&=&-i\tau \Gamma_R(\delta_{\tau,\ell}+\delta_{\bar\tau,\ell})\left[c_q\delta_{\nu,\tau}-s_qe^{-2i\nu\theta_q}\delta_{\nu,\bar\tau}\right],\\ \label{GammaRashba-cav}
\Gamma^{\rm cav}_S&=&i\tau \Gamma_S(\delta_{N\tau,\ell}-\delta_{N\bar\tau,\ell})\left[c_q\delta_{\nu,\tau}-s_qe^{-2i\nu\theta_q}\delta_{\nu,\bar\tau}\right],\label{GammaSkyrmion-cav}
\end{eqnarray}
where $\bar\tau=-\tau$ and $\bar\nu=-\nu$, and $\Gamma_\alpha=D_\alpha/\sqrt{\pi\omega_{{\bf k}=0}\epsilon_0L_z}$. The optical limit is easily recovered by setting ${\bf q}=0$ in $c_q$ and $s_q$. The value of $\gamma_0$ in a cavity is strongly increased by photon confinement with respect to the optical case. 

An additional dependence on ${\bf q}$ can arise from the exciton wave function via a Berry curvature dependent correction of the direct interaction term (see Appendix \ref{app:BSE}).

\section{Cavity exciton-polaritons}
\label{Sec:polaritons}

\begin{figure}[t]
\includegraphics[width=0.45\textwidth]{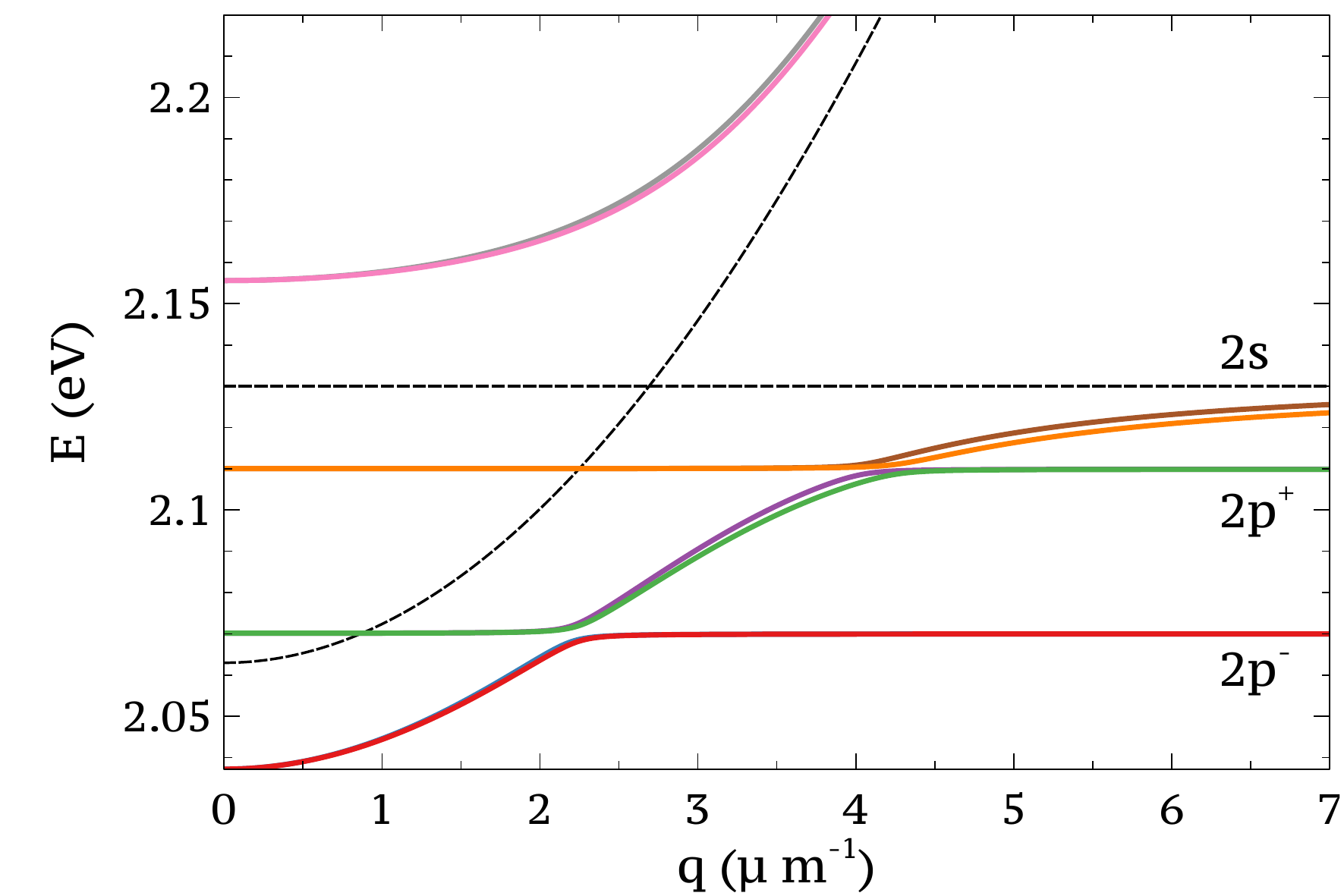}
\caption{Polartion bands resulting from the light-matter coupling described by the Hamiltonian Eq.~(\ref{PolHam}). Bare exciton energies are taken from Ref.~[\onlinecite{wu2015exciton}], $\epsilon_{2s}=2.13$ eV, $\epsilon^+_{2p^+}=2.11$ eV, $\epsilon^+_{2p^-}=2.07$ eV. 
\label{FigPolaritonBands}}
\end{figure}

We are now in the position to fully address the polaritonic bands of the exciton-photon system inside the cavity. As shown in the Appendix \ref{app:BSE}, the $2p$ exciton wavefunction acquires a $2s$ component at finite momentum. Although the correction is small we include it in the exciton Hamiltonian. The full Hamiltonian including $2s$, $2p$ states in both valleys and photons of both circular polarization reads
\begin{equation}\label{PolHam}
H=\left(\begin{array}{cccccccc}
\epsilon_{2s} & w_q & -w^ *_q & 0 & 0 & 0 & \gamma_q & \Gamma_q\\
w^ *_q & \epsilon^+_{2p^+} & 0 & 0 & 0 & 0 &\Gamma_c^* & \Gamma_s\\
-w_q & 0 & \epsilon^-_{2p^{-}} & 0 & 0 & 0 & -\Gamma_c^ * & -\Gamma_s\\
0 & 0 & 0 & \epsilon_{2s} & -w_q & w^ *_q & \Gamma_q^* & \gamma_q^ *\\
0 & 0 & 0 & -w^ *_q & \epsilon^-_{2p^-} & 0 & -\Gamma_s^ * & -\Gamma_c^ *\\
0 & 0 & 0 & w_q & 0 & \epsilon^-_{2p^{+}} & \Gamma_s^ * & \Gamma_c^ *\\
\gamma_q^ * & \Gamma_c & -\Gamma_c & \Gamma_q & -\Gamma_s & \Gamma_s & \omega_+ & 0\\
\Gamma_q^* & \Gamma_s^ * & -\Gamma_s^ * & \gamma_q & -\Gamma_c & \Gamma_{c} & 0 & \omega_-\\
\end{array}\right).
\end{equation}
where $w_q=\Delta_{sp}q a_B e^ {-i\theta_q}$ is the $2p$-$2s$ finite momentum coupliong, $\Gamma_c=i\Gamma_S c_q$, $\Gamma_s=i\Gamma_Ss_q e^ {-2i\theta_q}$ are cavity modified Skyrmion-mediated couplings, and $\gamma_q=i\gamma_0c_q$, $\Gamma_q=i\gamma_0 s_qe^ {2i\theta_q}$ are the couplings between photons and $2s$ excitons. The excitons energies are $\epsilon_{2s}=E_{2}+\Delta_{sp}$, $\epsilon^ \tau_{2p^ \pm}=E_2\pm \tau\Delta_{p}$, with $E_2$ the binding energy of $n=2$ states of the $V_q=2\pi e^ 2/(q(1+r_0q))$ screened potential, and the photon dispersion in the cavity is $\omega_\pm=c\sqrt{k_z^2+q^2}$, with $k_z=\pi/L_z$,.

In Fig.~\ref{FigPolaritonBands} we plot the resulting polaritons bands. We choose bare exciton energies for MoS$_2$ as in Ref.~[\onlinecite{wu2015exciton}], featuring splittings $\Delta_p=40$ meV and $\Delta_{sp}=40$ meV. The cavity width has been set to $L_z=0.3$ $\mu$m and the coupling $\gamma_0=(2/3)\times0.73$ meV, where the factor 2/3 accounts for the increased Bohr radius of $2s$ excitons. The factor $JV_Z/(\delta_c\delta_v)=0.05$ has been chosen, corresponding roughly to an in-plane magnetic field $B_x=5$ T and an exchange $J=50$ meV. 

The $2s$ excitons and the photons strongly couple and produce upper and lower polariton branches, with a Rabi frequency $\Omega_R=0.97$ eV. The polariton branches are slightly split by cavity effects\cite{gutierrez2018polariton}. In addition, we see that the Skyrmion-induced coupling produces small avoided crossings between the $2p$ states and the lower polariton branch

\begin{figure}[t]
\includegraphics[width=0.35\textwidth]{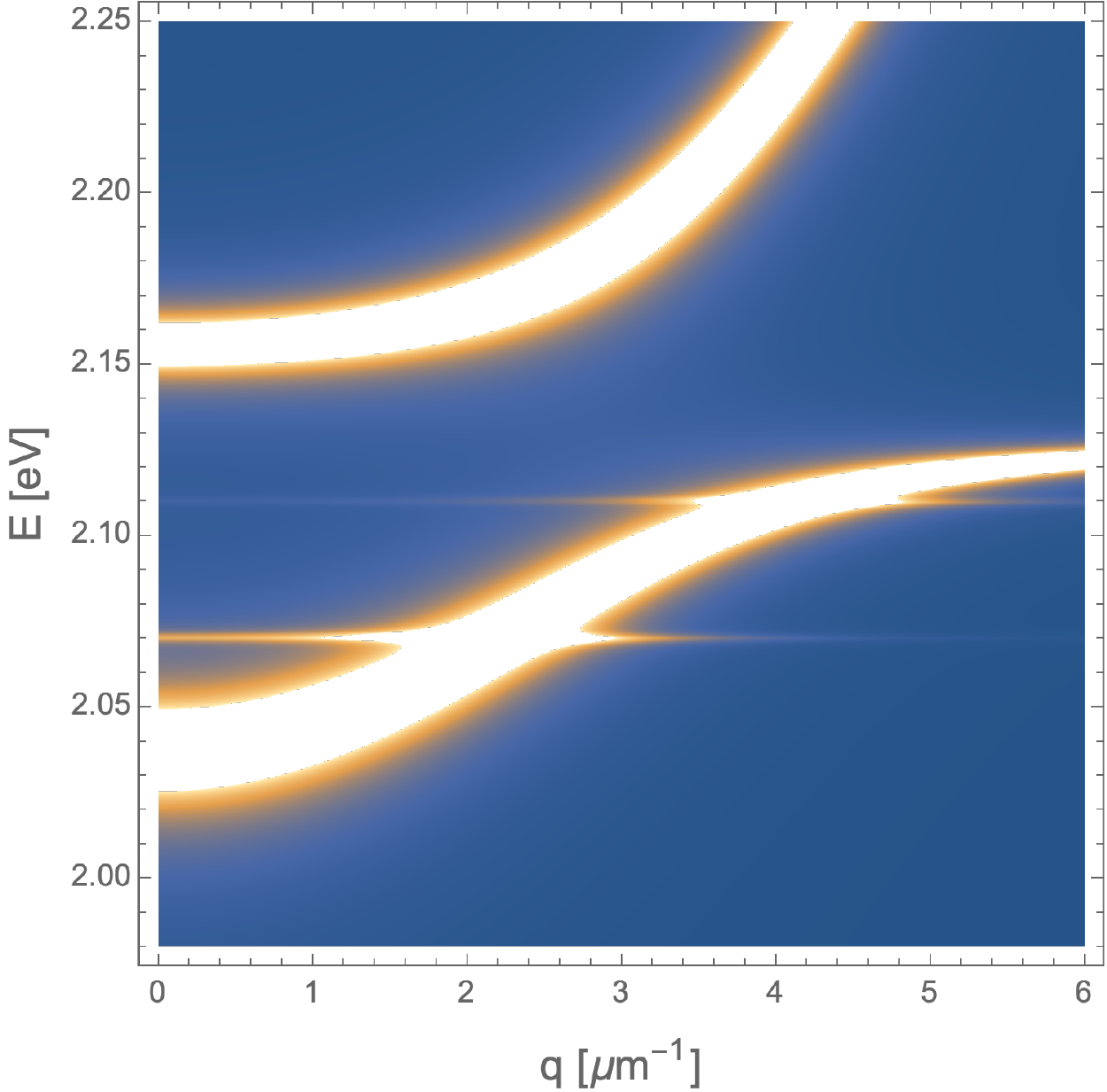}
\caption{Photon spectral function showing the photonic component of the polaritons. A photon loss of $\eta=1$ meV has been choosen. 
\label{FigPolariton}}
\end{figure}

In Fig.~\ref{FigPolariton} we plot the photon spectral function, that shows the photonic component of the polariton branches. A very high quality cavity has been assumed, characterized by 1 meV photon loss rate. Remarkably, beside a strong signal coming from the upper and lower polariton branches, we see the brightening of the $2p$ states, that couple to light via the matrix element Eq.~(\ref{GammaSkyrmion}). This result shows that, although the coupling mediated by an exciton is small and requires embedment of a Skyrmion in the TMD substrate, the strong light-matter coupling possible in a cavity and its high quality factor enable brightening of $2p$ states and measurement of the Berry curvature induced fine splittings.

\section{Conclusions}
\label{Sec:conclusions}

In this work we study how to externally engineer finite selection rules for coupling photon to $2p$ exciton by playing with the fundamental topological properties of the electronic carriers, with particular attention to the possibility to probe the predicted $2p^\pm$ exciton splitting. We show how the presence of a Rashba spin-orbit interaction or a Skyrmion in the underlaying substrate, together with the application of an in-plane magnetic field, can switch on a coupling between photons and all four $2p$ exciton states. For the case of the Rashba field, the winding imparted to the spin degree of freedom by the spin-orbit coupling mediates a unit of angular momentum at a cost of a spin-flip process, that has to be compensated by the in-plane magnetic field. For the case of a Skyrmion, its topological charge mediates the required unit of angular momentum. An analogous mechanism applies to unlike-spin excitons, for which the spin-flip process accompanying the transfer of angular momentum does not need to be compensated and directly provides the correct selection rules. 

Thanks to their stronger exchange field, Skyrmions provide a much stronger matrix element than Rashba SOI. Clearly, an experimental realization requires the embedding of a Skyrmion in the substrate, that is not expected to be a simple task. In order to amplify the signal, we suggest to embed the monolayer TMD in an optical cavity. The resulting finite momentum selection rules also describe the matrix element at the origin of cavity-polaritons. 

We believe that the results presented can shed light on the role of topological invariants in the coupling of electronic systems to photonic degrees of freedom and open the way to the engineering and control of topological light-matter coupling in cavity systems and topological polaritonics.

\section{Acknowledgments}

I acknowledge very useful discussions with R. Frisenda and P. Perna and I am particularly grateful to E. Ercolessi, for the hospitality at the Department of Physics of the University of Bologna, and to F. Guinea, for financial support through  funding from the Comunidad de Madrid through the grant MAD2D-CM, S2013/MIT-3007 and from the European Commission under the Graphene Flagship, contract CNECTICT-604391. I also acknowledge the European Commission for funding through the MCSA Global Fellowship grant TOPOCIRCUS-841894.

\appendix

\section{Bethe-Salpeter equation}
\label{app:BSE}

The full electronic Hamiltonian is written as ${\cal H}={\cal H}_0+{\cal V}$ with ${\cal H}_0=\sum_{\bf k}\epsilon^v_{\bf k}c^\dag_{v,{\bf k}}c_{v,{\bf k}}+\sum_{\bf k}\epsilon^c_{\bf k}c^\dag_{c,{\bf k}}c_{c,{\bf k}}$ and ${\cal V}=\frac{1}{2}\int d{\bf r}d{\bf r}'\hat{\psi}^\dag({\bf r})\hat{\psi}^\dag({\bf r}')V(|{\bf r}-{\bf r}'|)\hat{\psi}({\bf r}')\hat{\psi}({\bf r})$ where field operators are defined as $\hat{\psi}({\bf r})=\frac{1}{\sqrt S}\sum_{n,\tau,{\bf k}}u^{\tau,n}_{\bf k}({\bf r}) e^{i{\bf k}\cdot{\bf r}}c_{n,\tau,{\bf k}}$, with $S$ the sample area. Applying the full Hamiltonian onto exciton states Eq.~(\ref{ExcitonPhi}) and projecting it onto two-particle states we find the Bethe-Salpeter eigenvalue equation
\begin{eqnarray}
(\Delta_{{\bf k},{\bf q}}-E^{\ell,\tau}_{\bf q})\Phi^{\ell,\tau}_{{\bf k},{\bf q}}&=&\sum_{{\bf k}',\tau'}\left[D^{\tau}_{{\bf k},{\bf k}',{\bf q}}-X^{\tau,\tau'}_{{\bf k},{\bf k}',{\bf q}}\right]\Phi^{\ell,\tau'}_{{\bf k}',{\bf q}}
\end{eqnarray}
where $\Delta_{{\bf k},{\bf q}}=\epsilon^{c,\tau}_{{\bf k}+{\bf q}/2}-\epsilon^{v,\tau}_{{\bf k}-{\bf q}/2}$ and the direct and the exchange terms in $D$ and $X$ given by 
\begin{eqnarray}
D^{\tau}_{{\bf k},{\bf k}',{\bf q}}&=&V_{{\bf k}-{\bf k}'}s^{c\tau,c\tau}_{{\bf k}_+,{\bf k}'_+}s^{v\tau,v\tau}_{{\bf k}'_-,{\bf k}_-}\\
X^{\tau,\tau'}_{{\bf k},{\bf k}',{\bf q}}&=&V_{\bf q}s^{c\tau,v\tau}_{{\bf k}_+,{\bf k}_-}s^{v\tau',c\tau'}_{{\bf k}'_-,{\bf k}'_+}
\end{eqnarray}
with $s^{n\tau ,m\tau'}_{{\bf k},{\bf k}'}=\langle u^{n,\tau}_{\bf k}|u^{m,\tau'}_{{\bf k}'}\rangle$, $V_{\bf q}=\frac{1}{S}\int d{\bf r}V({\bf r})e^{i{\bf q}\cdot{\bf r}}$ and ${\bf k}_\pm={\bf k}\pm {\bf q}/2$. Following Ref.~[\onlinecite{srivastava2015signatures}] we notice that the overlaps $s^{n\tau ,m\tau'}_{{\bf k},{\bf k}'}$ contain informations about the wavefuction  coming from the two-band model Eq.~(\ref{Heff}). We then approximate the overlap matrix elements as
\begin{equation}
s^{c\tau ,c\tau'}_{{\bf k},{\bf k}'}=\delta_{\tau,\tau'}\left[1+i\tau\frac{\Omega_0}{4}({\bf k}'\times{\bf k})_z\right],
\end{equation}
where $\Omega_0$ is the Berry curvature of the conduction band electrons introduced in Sec.~\ref{Sec:ExcitonModel}. We separate the direct term in three contributions $D^\tau_{{\bf k},{\bf k}'}({\bf q})=V_{{\bf k}-{\bf k}'}+D^{(1)}_{{\bf k},{\bf k}'}+D^{(2)}_{{\bf k}-{\bf k}',{\bf q}}$. The first term is the screened quasi 2D Coulomb interaction that gives rise to the principal quantum number $n$ exciton spectrum featuring strong deviation from the $(n-1/2)^{-1}$. The second term $D^{(1)}_{{\bf k},{\bf k}'}$ was shown to be at the origin of non-hydrogenic effects in TMDCs \cite{srivastava2015signatures,wu2015exciton}. In particular, its symmetric part produces a splitting between $2s$ and $2p$ states and its antisymmetric part produces a splitting between $p^\pm$ states.  The last term is given by
\begin{equation}
D^{(2)}_{{\bf k}-{\bf k}',{\bf q}}=i\tau\frac{\Omega_0}{4}({\bf q}\times({\bf k}'-{\bf k}))_z V_{{\bf k}-{\bf k}'}.
\end{equation}

\begin{figure}[t]
\includegraphics[width=0.4\textwidth]{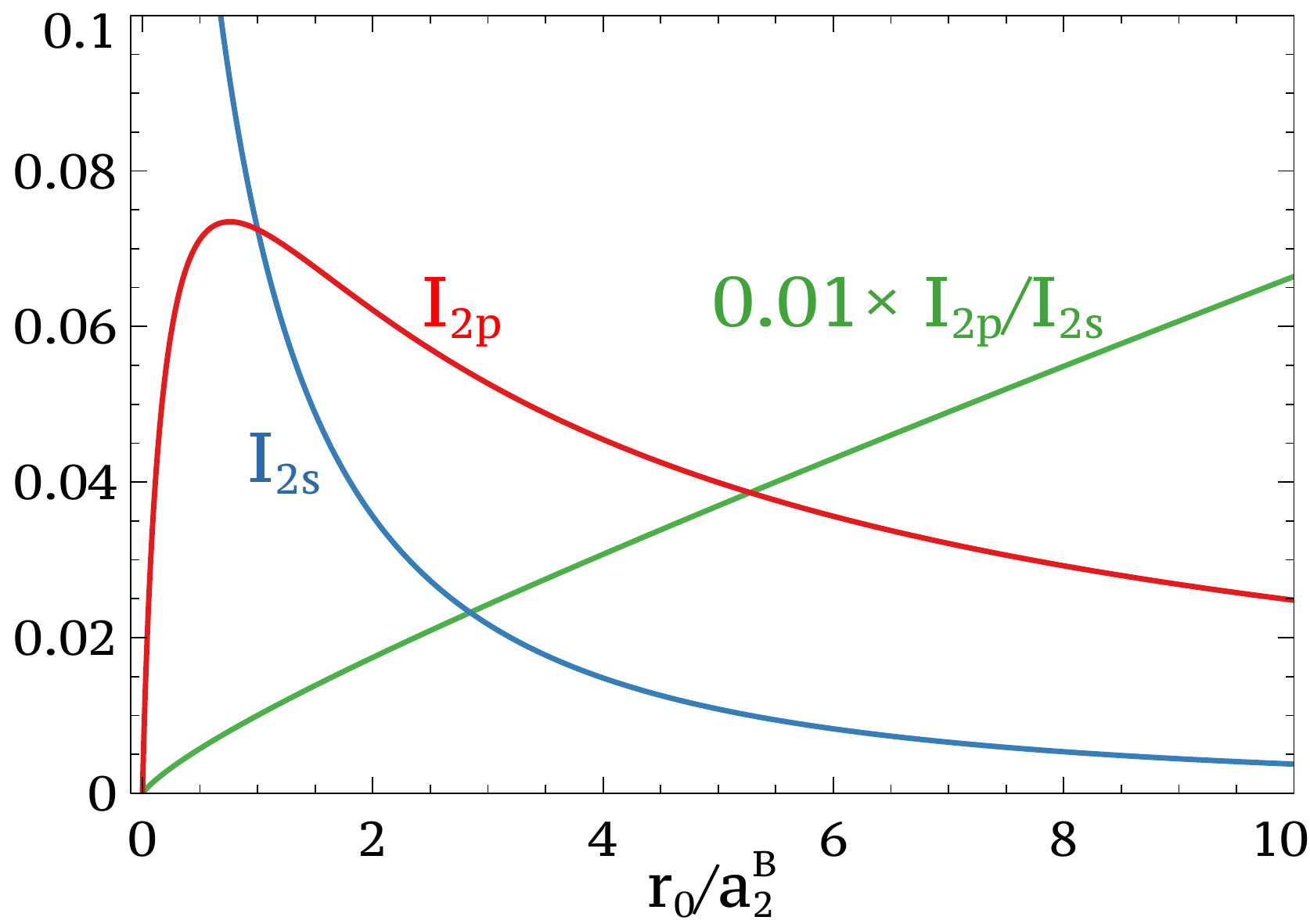}
\caption{Dependence of $I_{2p}$ and $I_{2s}$  as a function of $r_0/a^B_2$
\label{FigInteg}}
\end{figure}

This term is novel and it brings the ${\bf q}$-dependence arising due to a finite Berry curvature. The latter can be seen as a flux produced by the relative motion and center of mass motion of the particle hole pair and it represents a completely novel result. Interestingly, it  couples $2p$ states to $s$ states. For a pure Coulomb potential $V_k\propto 1/k$ and associated 2D hydrogenic exciton wavefunctions the coupling between $2s$ and $2p$ states turns out to be zero. However, for the screened potential $V_q=2\pi e^2/(q(1+r_0q))$ the matrix element can be reduced to 
\begin{equation}
w^\tau_{20;2\ell}=\tau\Omega_0qe^{i\ell \theta}\frac{\sqrt{\pi} e^2}{8\varepsilon (a^B_2)^2}I_{2p}(r_0/a^B_2)
\end{equation}
with $I_{2p}(z)=\int_0^\infty dx\frac{x^2(3-2x^2)}{(1+zx)(1+x^2)^{7/2}}$ and $a^B_2=3a_{\rm B}/2$ and $J_1$ a Bessel function of the first kind. The function $I_{2p}$ is plotted in Fig.~\ref{FigInteg} and we see that it has a maximum when the thickness of the material is on order of the exciton radius $a_{\rm B}\sim 1~{\rm nm}$. The expression for $w^\tau_{20;2\ell}$ has to be compared to the expression for the $2s$-$2p$ splitting generated by $D^ {(1)}$
\begin{equation}
\Delta_{sp}=\Omega_0\frac{e^2}{\varepsilon (a^B_2)^3}I_{2s}(\beta_2r_0).
\end{equation}
with $I_{2s}(z)=\int_0^\infty dx\frac{x^2(3x^2-2)}{(1+zx)(1+x^2)^{7/2}}$, that is shown in Fig.~Fig.~\ref{FigInteg}.  Within the framework of the approximation used we find 
\begin{equation}
\frac{w^\tau_{20;2\ell}}{\Delta_{sp}}=a_B\frac{3\sqrt{\pi}}{16}\frac{I_{2p}(r_0/a^B_2)}{I_{2s}(r_0/a^B_2)},
\end{equation}
In Fig.~\ref{FigInteg} we show the ratio $I_{2p}/I_{2s}$ and we conclude that $\delta\simeq a_B$.

Finally, the exchange term is zero for ${\bf q}=0$ due to the orthogonality of the valence and conduction band eigenstates at the same ${\bf k}$. Introducing ${\bf P}_{\tau,{\bf k}}=\langle\nabla_{\bf k}u^{v,\tau}_{{\bf k}}|u^{c,\tau}_{{\bf k}}\rangle$ the exchange interaction reads
\begin{equation}\label{MEexchange}
X^{\tau,\tau'}_{{\bf k},{\bf k}',{\bf q}}=V_{\bf q}\left({\bf q}\cdot{\bf P}_{\tau,{\bf k}}\right)^*{\bf q}\cdot{\bf P}_{\tau',{\bf k}'}.
\end{equation}
To address the matrix elements $\langle\Phi_{n,\ell,\tau}|X_{{\bf q}}|\Phi_{n',\ell',\tau'}\rangle$ a calculation in the direct space ${\bf r}$ is easier and it gives
\begin{equation}\label{MEx}
\sum_{{\bf k}}{\bf q}\cdot{\bf P}_{\tau,{\bf k}}\Phi_{n,\ell}({\bf k})\propto q (e^{i\theta_q}\delta_{\ell,\tau-1}+e^{-i\theta_q}\delta_{\ell,\tau+1}),
\end{equation}
with $\theta_q={\rm arg}(q_x+iq_y)$. Therefore, this term is non-zero only for $s$-wave and $d$-wave excitons and couples the two valleys. Its effect has been studied in Refs.~[\onlinecite{yu2014dirac,wu2015exciton,qiu2015nonanalyticity}] and it produces a finite momentum coupling between excitons in different valleys through a nonanalytic term proportional to $qe^{2i\theta}$.

\bibliography{polaritons}

\begin{thebibliography}{54}%
\makeatletter
\providecommand \@ifxundefined [1]{%
 \@ifx{#1\undefined}
}%
\providecommand \@ifnum [1]{%
 \ifnum #1\expandafter \@firstoftwo
 \else \expandafter \@secondoftwo
 \fi
}%
\providecommand \@ifx [1]{%
 \ifx #1\expandafter \@firstoftwo
 \else \expandafter \@secondoftwo
 \fi
}%
\providecommand \natexlab [1]{#1}%
\providecommand \enquote  [1]{``#1''}%
\providecommand \bibnamefont  [1]{#1}%
\providecommand \bibfnamefont [1]{#1}%
\providecommand \citenamefont [1]{#1}%
\providecommand \href@noop [0]{\@secondoftwo}%
\providecommand \href [0]{\begingroup \@sanitize@url \@href}%
\providecommand \@href[1]{\@@startlink{#1}\@@href}%
\providecommand \@@href[1]{\endgroup#1\@@endlink}%
\providecommand \@sanitize@url [0]{\catcode `\\12\catcode `\$12\catcode
  `\&12\catcode `\#12\catcode `\^12\catcode `\_12\catcode `\%12\relax}%
\providecommand \@@startlink[1]{}%
\providecommand \@@endlink[0]{}%
\providecommand \url  [0]{\begingroup\@sanitize@url \@url }%
\providecommand \@url [1]{\endgroup\@href {#1}{\urlprefix }}%
\providecommand \urlprefix  [0]{URL }%
\providecommand \Eprint [0]{\href }%
\providecommand \doibase [0]{http://dx.doi.org/}%
\providecommand \selectlanguage [0]{\@gobble}%
\providecommand \bibinfo  [0]{\@secondoftwo}%
\providecommand \bibfield  [0]{\@secondoftwo}%
\providecommand \translation [1]{[#1]}%
\providecommand \BibitemOpen [0]{}%
\providecommand \bibitemStop [0]{}%
\providecommand \bibitemNoStop [0]{.\EOS\space}%
\providecommand \EOS [0]{\spacefactor3000\relax}%
\providecommand \BibitemShut  [1]{\csname bibitem#1\endcsname}%
\let\auto@bib@innerbib\@empty
\bibitem [{\citenamefont {Cao}\ \emph {et~al.}(2018)\citenamefont {Cao},
  \citenamefont {Wu},\ and\ \citenamefont {Louie}}]{cao2018unifying}%
  \BibitemOpen
  \bibfield  {author} {\bibinfo {author} {\bibfnamefont {T.}~\bibnamefont
  {Cao}}, \bibinfo {author} {\bibfnamefont {M.}~\bibnamefont {Wu}}, \ and\
  \bibinfo {author} {\bibfnamefont {S.~G.}\ \bibnamefont {Louie}},\ }\href@noop
  {} {\bibfield  {journal} {\bibinfo  {journal} {Phys. Rev. Lett.}\ }\textbf
  {\bibinfo {volume} {120}},\ \bibinfo {pages} {087402} (\bibinfo {year}
  {2018})}\BibitemShut {NoStop}%
\bibitem [{\citenamefont {Zhang}\ \emph {et~al.}(2018)\citenamefont {Zhang},
  \citenamefont {Shan},\ and\ \citenamefont {Xiao}}]{zhang2018optical}%
  \BibitemOpen
  \bibfield  {author} {\bibinfo {author} {\bibfnamefont {X.}~\bibnamefont
  {Zhang}}, \bibinfo {author} {\bibfnamefont {W.-Y.}\ \bibnamefont {Shan}}, \
  and\ \bibinfo {author} {\bibfnamefont {D.}~\bibnamefont {Xiao}},\ }\href@noop
  {} {\bibfield  {journal} {\bibinfo  {journal} {Phys. Rev. Lett.}\ }\textbf
  {\bibinfo {volume} {120}},\ \bibinfo {pages} {077401} (\bibinfo {year}
  {2018})}\BibitemShut {NoStop}%
\bibitem [{\citenamefont {Rold{\'{a}}n}\ \emph {et~al.}(2017)\citenamefont
  {Rold{\'{a}}n}, \citenamefont {Chirolli}, \citenamefont {Prada},
  \citenamefont {Silva-Guill{\'{e}}n}, \citenamefont {San-Jose},\ and\
  \citenamefont {Guinea}}]{roldan2017theory}%
  \BibitemOpen
  \bibfield  {author} {\bibinfo {author} {\bibfnamefont {R.}~\bibnamefont
  {Rold{\'{a}}n}}, \bibinfo {author} {\bibfnamefont {L.}~\bibnamefont
  {Chirolli}}, \bibinfo {author} {\bibfnamefont {E.}~\bibnamefont {Prada}},
  \bibinfo {author} {\bibfnamefont {J.~A.}\ \bibnamefont
  {Silva-Guill{\'{e}}n}}, \bibinfo {author} {\bibfnamefont {P.}~\bibnamefont
  {San-Jose}}, \ and\ \bibinfo {author} {\bibfnamefont {F.}~\bibnamefont
  {Guinea}},\ }\href {\doibase 10.1039/C7CS00210F} {\bibfield  {journal}
  {\bibinfo  {journal} {Chemical Society Reviews}\ }\textbf {\bibinfo {volume}
  {46}},\ \bibinfo {pages} {4387} (\bibinfo {year} {2017})}\BibitemShut
  {NoStop}%
\bibitem [{\citenamefont {Ugeda}\ \emph {et~al.}(2014)\citenamefont {Ugeda},
  \citenamefont {Bradley}, \citenamefont {Shi}, \citenamefont {da~Jornada},
  \citenamefont {Zhang}, \citenamefont {Qiu}, \citenamefont {Ruan},
  \citenamefont {Mo}, \citenamefont {Hussain}, \citenamefont {Shen},
  \citenamefont {Wang}, \citenamefont {Louie},\ and\ \citenamefont
  {Crommie}}]{ugeda2014giant}%
  \BibitemOpen
  \bibfield  {author} {\bibinfo {author} {\bibfnamefont {M.~M.}\ \bibnamefont
  {Ugeda}}, \bibinfo {author} {\bibfnamefont {A.~J.}\ \bibnamefont {Bradley}},
  \bibinfo {author} {\bibfnamefont {S.-F.}\ \bibnamefont {Shi}}, \bibinfo
  {author} {\bibfnamefont {F.~H.}\ \bibnamefont {da~Jornada}}, \bibinfo
  {author} {\bibfnamefont {Y.}~\bibnamefont {Zhang}}, \bibinfo {author}
  {\bibfnamefont {D.~Y.}\ \bibnamefont {Qiu}}, \bibinfo {author} {\bibfnamefont
  {W.}~\bibnamefont {Ruan}}, \bibinfo {author} {\bibfnamefont {S.-K.}\
  \bibnamefont {Mo}}, \bibinfo {author} {\bibfnamefont {Z.}~\bibnamefont
  {Hussain}}, \bibinfo {author} {\bibfnamefont {Z.-X.}\ \bibnamefont {Shen}},
  \bibinfo {author} {\bibfnamefont {F.}~\bibnamefont {Wang}}, \bibinfo {author}
  {\bibfnamefont {S.~G.}\ \bibnamefont {Louie}}, \ and\ \bibinfo {author}
  {\bibfnamefont {M.~F.}\ \bibnamefont {Crommie}},\ }\href@noop {} {\bibfield
  {journal} {\bibinfo  {journal} {Nature Materials}\ }\textbf {\bibinfo
  {volume} {13}},\ \bibinfo {pages} {1091 EP } (\bibinfo {year}
  {2014})}\BibitemShut {NoStop}%
\bibitem [{\citenamefont {Mueller}\ and\ \citenamefont
  {Malic}(2018)}]{MuellerMalic}%
  \BibitemOpen
  \bibfield  {author} {\bibinfo {author} {\bibfnamefont {T.}~\bibnamefont
  {Mueller}}\ and\ \bibinfo {author} {\bibfnamefont {E.}~\bibnamefont
  {Malic}},\ }\href {\doibase 10.1038/s41699-018-0074-2} {\bibfield  {journal}
  {\bibinfo  {journal} {npj 2D Materials and Applications}\ }\textbf {\bibinfo
  {volume} {2}},\ \bibinfo {pages} {29} (\bibinfo {year} {2018})}\BibitemShut
  {NoStop}%
\bibitem [{\citenamefont {Yu}\ \emph {et~al.}(2015)\citenamefont {Yu},
  \citenamefont {Cui}, \citenamefont {Xu},\ and\ \citenamefont
  {Yao}}]{yu2015valley}%
  \BibitemOpen
  \bibfield  {author} {\bibinfo {author} {\bibfnamefont {H.}~\bibnamefont
  {Yu}}, \bibinfo {author} {\bibfnamefont {X.}~\bibnamefont {Cui}}, \bibinfo
  {author} {\bibfnamefont {X.}~\bibnamefont {Xu}}, \ and\ \bibinfo {author}
  {\bibfnamefont {W.}~\bibnamefont {Yao}},\ }\href@noop {} {\bibfield
  {journal} {\bibinfo  {journal} {National Science Review}\ }\textbf {\bibinfo
  {volume} {2}},\ \bibinfo {pages} {57} (\bibinfo {year} {2015})}\BibitemShut
  {NoStop}%
\bibitem [{\citenamefont {Manzeli}\ \emph {et~al.}(2017)\citenamefont
  {Manzeli}, \citenamefont {Ovchinnikov}, \citenamefont {Pasquier},
  \citenamefont {Yazyev},\ and\ \citenamefont {Kis}}]{manzeli20172d}%
  \BibitemOpen
  \bibfield  {author} {\bibinfo {author} {\bibfnamefont {S.}~\bibnamefont
  {Manzeli}}, \bibinfo {author} {\bibfnamefont {D.}~\bibnamefont
  {Ovchinnikov}}, \bibinfo {author} {\bibfnamefont {D.}~\bibnamefont
  {Pasquier}}, \bibinfo {author} {\bibfnamefont {O.~V.}\ \bibnamefont
  {Yazyev}}, \ and\ \bibinfo {author} {\bibfnamefont {A.}~\bibnamefont {Kis}},\
  }\href {https://doi.org/10.1038/natrevmats.2017.33} {\bibfield  {journal}
  {\bibinfo  {journal} {Nature Reviews Materials}\ }\textbf {\bibinfo {volume}
  {2}},\ \bibinfo {pages} {17033 EP } (\bibinfo {year} {2017})}\BibitemShut
  {NoStop}%
\bibitem [{\citenamefont {Wang}\ \emph {et~al.}(2018)\citenamefont {Wang},
  \citenamefont {Chernikov}, \citenamefont {Glazov}, \citenamefont {Heinz},
  \citenamefont {Marie}, \citenamefont {Amand},\ and\ \citenamefont
  {Urbaszek}}]{wang2018colloquium}%
  \BibitemOpen
  \bibfield  {author} {\bibinfo {author} {\bibfnamefont {G.}~\bibnamefont
  {Wang}}, \bibinfo {author} {\bibfnamefont {A.}~\bibnamefont {Chernikov}},
  \bibinfo {author} {\bibfnamefont {M.~M.}\ \bibnamefont {Glazov}}, \bibinfo
  {author} {\bibfnamefont {T.~F.}\ \bibnamefont {Heinz}}, \bibinfo {author}
  {\bibfnamefont {X.}~\bibnamefont {Marie}}, \bibinfo {author} {\bibfnamefont
  {T.}~\bibnamefont {Amand}}, \ and\ \bibinfo {author} {\bibfnamefont
  {B.}~\bibnamefont {Urbaszek}},\ }\href {\doibase
  10.1103/RevModPhys.90.021001} {\bibfield  {journal} {\bibinfo  {journal}
  {Rev. Mod. Phys.}\ }\textbf {\bibinfo {volume} {90}},\ \bibinfo {pages}
  {021001} (\bibinfo {year} {2018})}\BibitemShut {NoStop}%
\bibitem [{\citenamefont {Hsu}\ \emph {et~al.}(2019)\citenamefont {Hsu},
  \citenamefont {Frisenda}, \citenamefont {Schmidt}, \citenamefont {Arora},
  \citenamefont {de~Vasconcellos}, \citenamefont {Bratschitsch}, \citenamefont
  {van~der Zant},\ and\ \citenamefont
  {Castellanos-Gomez}}]{hsu2019thickness-dependent}%
  \BibitemOpen
  \bibfield  {author} {\bibinfo {author} {\bibfnamefont {C.}~\bibnamefont
  {Hsu}}, \bibinfo {author} {\bibfnamefont {R.}~\bibnamefont {Frisenda}},
  \bibinfo {author} {\bibfnamefont {R.}~\bibnamefont {Schmidt}}, \bibinfo
  {author} {\bibfnamefont {A.}~\bibnamefont {Arora}}, \bibinfo {author}
  {\bibfnamefont {S.~M.}\ \bibnamefont {de~Vasconcellos}}, \bibinfo {author}
  {\bibfnamefont {R.}~\bibnamefont {Bratschitsch}}, \bibinfo {author}
  {\bibfnamefont {H.~S.~J.}\ \bibnamefont {van~der Zant}}, \ and\ \bibinfo
  {author} {\bibfnamefont {A.}~\bibnamefont {Castellanos-Gomez}},\ }\href
  {\doibase 10.1002/adom.201900239} {\bibfield  {journal} {\bibinfo  {journal}
  {Advanced Optical Materials}\ }\textbf {\bibinfo {volume} {7}},\ \bibinfo
  {pages} {1900239} (\bibinfo {year} {2019})}\BibitemShut {NoStop}%
\bibitem [{\citenamefont {He}\ \emph {et~al.}(2014)\citenamefont {He},
  \citenamefont {Kumar}, \citenamefont {Zhao}, \citenamefont {Wang},
  \citenamefont {Mak}, \citenamefont {Zhao},\ and\ \citenamefont
  {Shan}}]{he2014tightly}%
  \BibitemOpen
  \bibfield  {author} {\bibinfo {author} {\bibfnamefont {K.}~\bibnamefont
  {He}}, \bibinfo {author} {\bibfnamefont {N.}~\bibnamefont {Kumar}}, \bibinfo
  {author} {\bibfnamefont {L.}~\bibnamefont {Zhao}}, \bibinfo {author}
  {\bibfnamefont {Z.}~\bibnamefont {Wang}}, \bibinfo {author} {\bibfnamefont
  {K.~F.}\ \bibnamefont {Mak}}, \bibinfo {author} {\bibfnamefont
  {H.}~\bibnamefont {Zhao}}, \ and\ \bibinfo {author} {\bibfnamefont
  {J.}~\bibnamefont {Shan}},\ }\href {\doibase 10.1103/PhysRevLett.113.026803}
  {\bibfield  {journal} {\bibinfo  {journal} {Phys. Rev. Lett.}\ }\textbf
  {\bibinfo {volume} {113}},\ \bibinfo {pages} {026803} (\bibinfo {year}
  {2014})}\BibitemShut {NoStop}%
\bibitem [{\citenamefont {Chernikov}\ \emph {et~al.}(2014)\citenamefont
  {Chernikov}, \citenamefont {Berkelbach}, \citenamefont {Hill}, \citenamefont
  {Rigosi}, \citenamefont {Li}, \citenamefont {Aslan}, \citenamefont
  {Reichman}, \citenamefont {Hybertsen},\ and\ \citenamefont
  {Heinz}}]{chernikov2014exciton}%
  \BibitemOpen
  \bibfield  {author} {\bibinfo {author} {\bibfnamefont {A.}~\bibnamefont
  {Chernikov}}, \bibinfo {author} {\bibfnamefont {T.~C.}\ \bibnamefont
  {Berkelbach}}, \bibinfo {author} {\bibfnamefont {H.~M.}\ \bibnamefont
  {Hill}}, \bibinfo {author} {\bibfnamefont {A.}~\bibnamefont {Rigosi}},
  \bibinfo {author} {\bibfnamefont {Y.}~\bibnamefont {Li}}, \bibinfo {author}
  {\bibfnamefont {O.~B.}\ \bibnamefont {Aslan}}, \bibinfo {author}
  {\bibfnamefont {D.~R.}\ \bibnamefont {Reichman}}, \bibinfo {author}
  {\bibfnamefont {M.~S.}\ \bibnamefont {Hybertsen}}, \ and\ \bibinfo {author}
  {\bibfnamefont {T.~F.}\ \bibnamefont {Heinz}},\ }\href {\doibase
  10.1103/PhysRevLett.113.076802} {\bibfield  {journal} {\bibinfo  {journal}
  {Phys. Rev. Lett.}\ }\textbf {\bibinfo {volume} {113}},\ \bibinfo {pages}
  {076802} (\bibinfo {year} {2014})}\BibitemShut {NoStop}%
\bibitem [{\citenamefont {Liu}\ \emph {et~al.}(2019)\citenamefont {Liu},
  \citenamefont {van Baren}, \citenamefont {Taniguchi}, \citenamefont
  {Watanabe}, \citenamefont {Chang},\ and\ \citenamefont
  {Lui}}]{liu2019magnetophotoluminescence}%
  \BibitemOpen
  \bibfield  {author} {\bibinfo {author} {\bibfnamefont {E.}~\bibnamefont
  {Liu}}, \bibinfo {author} {\bibfnamefont {J.}~\bibnamefont {van Baren}},
  \bibinfo {author} {\bibfnamefont {T.}~\bibnamefont {Taniguchi}}, \bibinfo
  {author} {\bibfnamefont {K.}~\bibnamefont {Watanabe}}, \bibinfo {author}
  {\bibfnamefont {Y.-C.}\ \bibnamefont {Chang}}, \ and\ \bibinfo {author}
  {\bibfnamefont {C.~H.}\ \bibnamefont {Lui}},\ }\href {\doibase
  10.1103/PhysRevB.99.205420} {\bibfield  {journal} {\bibinfo  {journal} {Phys.
  Rev. B}\ }\textbf {\bibinfo {volume} {99}},\ \bibinfo {pages} {205420}
  (\bibinfo {year} {2019})}\BibitemShut {NoStop}%
\bibitem [{\citenamefont {Qiu}\ \emph {et~al.}(2015)\citenamefont {Qiu},
  \citenamefont {Cao},\ and\ \citenamefont {Louie}}]{qiu2015nonanalyticity}%
  \BibitemOpen
  \bibfield  {author} {\bibinfo {author} {\bibfnamefont {D.~Y.}\ \bibnamefont
  {Qiu}}, \bibinfo {author} {\bibfnamefont {T.}~\bibnamefont {Cao}}, \ and\
  \bibinfo {author} {\bibfnamefont {S.~G.}\ \bibnamefont {Louie}},\ }\href@noop
  {} {\bibfield  {journal} {\bibinfo  {journal} {Phys. Rev. Lett.}\ }\textbf
  {\bibinfo {volume} {115}},\ \bibinfo {pages} {176801} (\bibinfo {year}
  {2015})}\BibitemShut {NoStop}%
\bibitem [{\citenamefont {Wu}\ \emph {et~al.}(2015)\citenamefont {Wu},
  \citenamefont {Qu},\ and\ \citenamefont {MacDonald}}]{wu2015exciton}%
  \BibitemOpen
  \bibfield  {author} {\bibinfo {author} {\bibfnamefont {F.}~\bibnamefont
  {Wu}}, \bibinfo {author} {\bibfnamefont {F.}~\bibnamefont {Qu}}, \ and\
  \bibinfo {author} {\bibfnamefont {A.~H.}\ \bibnamefont {MacDonald}},\ }\href
  {\doibase 10.1103/PhysRevB.91.075310} {\bibfield  {journal} {\bibinfo
  {journal} {Physical Review B}\ }\textbf {\bibinfo {volume} {91}},\ \bibinfo
  {pages} {075310} (\bibinfo {year} {2015})}\BibitemShut {NoStop}%
\bibitem [{\citenamefont {Srivastava}\ and\ \citenamefont {Imamo{\u
  g}lu}(2015)}]{srivastava2015signatures}%
  \BibitemOpen
  \bibfield  {author} {\bibinfo {author} {\bibfnamefont {A.}~\bibnamefont
  {Srivastava}}\ and\ \bibinfo {author} {\bibfnamefont {A.}~\bibnamefont
  {Imamo{\u g}lu}},\ }\href {\doibase 10.1103/PhysRevLett.115.166802}
  {\bibfield  {journal} {\bibinfo  {journal} {Physical Review Letters}\
  }\textbf {\bibinfo {volume} {115}},\ \bibinfo {pages} {166802} (\bibinfo
  {year} {2015})}\BibitemShut {NoStop}%
\bibitem [{\citenamefont {Zhou}\ \emph {et~al.}(2015)\citenamefont {Zhou},
  \citenamefont {Shan}, \citenamefont {Yao},\ and\ \citenamefont
  {Xiao}}]{zhou2015berry}%
  \BibitemOpen
  \bibfield  {author} {\bibinfo {author} {\bibfnamefont {J.}~\bibnamefont
  {Zhou}}, \bibinfo {author} {\bibfnamefont {W.-Y.}\ \bibnamefont {Shan}},
  \bibinfo {author} {\bibfnamefont {W.}~\bibnamefont {Yao}}, \ and\ \bibinfo
  {author} {\bibfnamefont {D.}~\bibnamefont {Xiao}},\ }\href {\doibase
  10.1103/PhysRevLett.115.166803} {\bibfield  {journal} {\bibinfo  {journal}
  {Physical Review Letters}\ }\textbf {\bibinfo {volume} {115}},\ \bibinfo
  {pages} {166803} (\bibinfo {year} {2015})}\BibitemShut {NoStop}%
\bibitem [{\citenamefont {Ye}\ \emph {et~al.}(2014)\citenamefont {Ye},
  \citenamefont {Cao}, \citenamefont {O'Brien}, \citenamefont {Zhu},
  \citenamefont {Yin}, \citenamefont {Wang}, \citenamefont {Louie},\ and\
  \citenamefont {Zhang}}]{ye2014probing}%
  \BibitemOpen
  \bibfield  {author} {\bibinfo {author} {\bibfnamefont {Z.}~\bibnamefont
  {Ye}}, \bibinfo {author} {\bibfnamefont {T.}~\bibnamefont {Cao}}, \bibinfo
  {author} {\bibfnamefont {K.}~\bibnamefont {O'Brien}}, \bibinfo {author}
  {\bibfnamefont {H.}~\bibnamefont {Zhu}}, \bibinfo {author} {\bibfnamefont
  {X.}~\bibnamefont {Yin}}, \bibinfo {author} {\bibfnamefont {Y.}~\bibnamefont
  {Wang}}, \bibinfo {author} {\bibfnamefont {S.~G.}\ \bibnamefont {Louie}}, \
  and\ \bibinfo {author} {\bibfnamefont {X.}~\bibnamefont {Zhang}},\ }\href
  {https://doi.org/10.1038/nature13734} {\bibfield  {journal} {\bibinfo
  {journal} {Nature}\ }\textbf {\bibinfo {volume} {513}},\ \bibinfo {pages}
  {214 EP } (\bibinfo {year} {2014})}\BibitemShut {NoStop}%
\bibitem [{\citenamefont {Wang}\ \emph {et~al.}(2015)\citenamefont {Wang},
  \citenamefont {Marie}, \citenamefont {Gerber}, \citenamefont {Amand},
  \citenamefont {Lagarde}, \citenamefont {Bouet}, \citenamefont {Vidal},
  \citenamefont {Balocchi},\ and\ \citenamefont {Urbaszek}}]{wang2015giant}%
  \BibitemOpen
  \bibfield  {author} {\bibinfo {author} {\bibfnamefont {G.}~\bibnamefont
  {Wang}}, \bibinfo {author} {\bibfnamefont {X.}~\bibnamefont {Marie}},
  \bibinfo {author} {\bibfnamefont {I.}~\bibnamefont {Gerber}}, \bibinfo
  {author} {\bibfnamefont {T.}~\bibnamefont {Amand}}, \bibinfo {author}
  {\bibfnamefont {D.}~\bibnamefont {Lagarde}}, \bibinfo {author} {\bibfnamefont
  {L.}~\bibnamefont {Bouet}}, \bibinfo {author} {\bibfnamefont
  {M.}~\bibnamefont {Vidal}}, \bibinfo {author} {\bibfnamefont
  {A.}~\bibnamefont {Balocchi}}, \ and\ \bibinfo {author} {\bibfnamefont
  {B.}~\bibnamefont {Urbaszek}},\ }\href {\doibase
  10.1103/PhysRevLett.114.097403} {\bibfield  {journal} {\bibinfo  {journal}
  {Phys. Rev. Lett.}\ }\textbf {\bibinfo {volume} {114}},\ \bibinfo {pages}
  {097403} (\bibinfo {year} {2015})}\BibitemShut {NoStop}%
\bibitem [{\citenamefont {Karzig}\ \emph {et~al.}(2015)\citenamefont {Karzig},
  \citenamefont {Bardyn}, \citenamefont {Lindner},\ and\ \citenamefont
  {Refael}}]{karzig2015topological}%
  \BibitemOpen
  \bibfield  {author} {\bibinfo {author} {\bibfnamefont {T.}~\bibnamefont
  {Karzig}}, \bibinfo {author} {\bibfnamefont {C.-E.}\ \bibnamefont {Bardyn}},
  \bibinfo {author} {\bibfnamefont {N.~H.}\ \bibnamefont {Lindner}}, \ and\
  \bibinfo {author} {\bibfnamefont {G.}~\bibnamefont {Refael}},\ }\href
  {\doibase 10.1103/PhysRevX.5.031001} {\bibfield  {journal} {\bibinfo
  {journal} {Physical Review X}\ }\textbf {\bibinfo {volume} {5}},\ \bibinfo
  {pages} {031001} (\bibinfo {year} {2015})}\BibitemShut {NoStop}%
\bibitem [{\citenamefont {Guti\'errez-Rubio}\ \emph {et~al.}(2018)\citenamefont
  {Guti\'errez-Rubio}, \citenamefont {Chirolli}, \citenamefont
  {Mart\'{\i}n-Moreno}, \citenamefont {Garc\'{\i}a-Vidal},\ and\ \citenamefont
  {Guinea}}]{gutierrez2018polariton}%
  \BibitemOpen
  \bibfield  {author} {\bibinfo {author} {\bibfnamefont {A.}~\bibnamefont
  {Guti\'errez-Rubio}}, \bibinfo {author} {\bibfnamefont {L.}~\bibnamefont
  {Chirolli}}, \bibinfo {author} {\bibfnamefont {L.}~\bibnamefont
  {Mart\'{\i}n-Moreno}}, \bibinfo {author} {\bibfnamefont {F.~J.}\ \bibnamefont
  {Garc\'{\i}a-Vidal}}, \ and\ \bibinfo {author} {\bibfnamefont
  {F.}~\bibnamefont {Guinea}},\ }\href@noop {} {\bibfield  {journal} {\bibinfo
  {journal} {Phys. Rev. Lett.}\ }\textbf {\bibinfo {volume} {121}},\ \bibinfo
  {pages} {137402} (\bibinfo {year} {2018})}\BibitemShut {NoStop}%
\bibitem [{\citenamefont {Latini}\ \emph {et~al.}(2019)\citenamefont {Latini},
  \citenamefont {Ronca}, \citenamefont {De~Giovannini}, \citenamefont
  {H{\"u}bener},\ and\ \citenamefont {Rubio}}]{latini2019cavity}%
  \BibitemOpen
  \bibfield  {author} {\bibinfo {author} {\bibfnamefont {S.}~\bibnamefont
  {Latini}}, \bibinfo {author} {\bibfnamefont {E.}~\bibnamefont {Ronca}},
  \bibinfo {author} {\bibfnamefont {U.}~\bibnamefont {De~Giovannini}}, \bibinfo
  {author} {\bibfnamefont {H.}~\bibnamefont {H{\"u}bener}}, \ and\ \bibinfo
  {author} {\bibfnamefont {A.}~\bibnamefont {Rubio}},\ }\href@noop {}
  {\bibfield  {journal} {\bibinfo  {journal} {Nano Letters}\ }\textbf {\bibinfo
  {volume} {19}},\ \bibinfo {pages} {3473} (\bibinfo {year}
  {2019})}\BibitemShut {NoStop}%
\bibitem [{\citenamefont {Hasan}\ and\ \citenamefont
  {Kane}(2010)}]{hasan2010colloquium}%
  \BibitemOpen
  \bibfield  {author} {\bibinfo {author} {\bibfnamefont {M.~Z.}\ \bibnamefont
  {Hasan}}\ and\ \bibinfo {author} {\bibfnamefont {C.~L.}\ \bibnamefont
  {Kane}},\ }\href {\doibase 10.1103/RevModPhys.82.3045} {\bibfield  {journal}
  {\bibinfo  {journal} {Rev. Mod. Phys.}\ }\textbf {\bibinfo {volume} {82}},\
  \bibinfo {pages} {3045} (\bibinfo {year} {2010})}\BibitemShut {NoStop}%
\bibitem [{\citenamefont {Qi}\ and\ \citenamefont
  {Zhang}(2011)}]{qi2011topological}%
  \BibitemOpen
  \bibfield  {author} {\bibinfo {author} {\bibfnamefont {X.-L.}\ \bibnamefont
  {Qi}}\ and\ \bibinfo {author} {\bibfnamefont {S.-C.}\ \bibnamefont {Zhang}},\
  }\href {\doibase 10.1103/RevModPhys.83.1057} {\bibfield  {journal} {\bibinfo
  {journal} {Rev. Mod. Phys.}\ }\textbf {\bibinfo {volume} {83}},\ \bibinfo
  {pages} {1057} (\bibinfo {year} {2011})}\BibitemShut {NoStop}%
\bibitem [{\citenamefont {Ando}(2013)}]{ando2013topological}%
  \BibitemOpen
  \bibfield  {author} {\bibinfo {author} {\bibfnamefont {Y.}~\bibnamefont
  {Ando}},\ }\href@noop {} {\bibfield  {journal} {\bibinfo  {journal} {Journal
  of the Physical Society of Japan}\ }\textbf {\bibinfo {volume} {82}},\
  \bibinfo {pages} {102001} (\bibinfo {year} {2013})}\BibitemShut {NoStop}%
\bibitem [{\citenamefont {Korm\'anyos}\ \emph {et~al.}(2014)\citenamefont
  {Korm\'anyos}, \citenamefont {Z\'olyomi}, \citenamefont {Drummond},\ and\
  \citenamefont {Burkard}}]{kormanyos2014spin-orbit}%
  \BibitemOpen
  \bibfield  {author} {\bibinfo {author} {\bibfnamefont {A.}~\bibnamefont
  {Korm\'anyos}}, \bibinfo {author} {\bibfnamefont {V.}~\bibnamefont
  {Z\'olyomi}}, \bibinfo {author} {\bibfnamefont {N.~D.}\ \bibnamefont
  {Drummond}}, \ and\ \bibinfo {author} {\bibfnamefont {G.}~\bibnamefont
  {Burkard}},\ }\href {\doibase 10.1103/PhysRevX.4.011034} {\bibfield
  {journal} {\bibinfo  {journal} {Phys. Rev. X}\ }\textbf {\bibinfo {volume}
  {4}},\ \bibinfo {pages} {011034} (\bibinfo {year} {2014})}\BibitemShut
  {NoStop}%
\bibitem [{\citenamefont {Nagaosa}\ and\ \citenamefont
  {Tokura}(2013)}]{nagaosa2013topological}%
  \BibitemOpen
  \bibfield  {author} {\bibinfo {author} {\bibfnamefont {N.}~\bibnamefont
  {Nagaosa}}\ and\ \bibinfo {author} {\bibfnamefont {Y.}~\bibnamefont
  {Tokura}},\ }\href {https://doi.org/10.1038/nnano.2013.243} {\bibfield
  {journal} {\bibinfo  {journal} {Nature Nanotechnology}\ }\textbf {\bibinfo
  {volume} {8}},\ \bibinfo {pages} {899 EP } (\bibinfo {year}
  {2013})}\BibitemShut {NoStop}%
\bibitem [{\citenamefont {Fert}\ \emph {et~al.}(2017)\citenamefont {Fert},
  \citenamefont {Reyren},\ and\ \citenamefont {Cros}}]{fert2017magnetic}%
  \BibitemOpen
  \bibfield  {author} {\bibinfo {author} {\bibfnamefont {A.}~\bibnamefont
  {Fert}}, \bibinfo {author} {\bibfnamefont {N.}~\bibnamefont {Reyren}}, \ and\
  \bibinfo {author} {\bibfnamefont {V.}~\bibnamefont {Cros}},\ }\href
  {https://doi.org/10.1038/natrevmats.2017.31} {\bibfield  {journal} {\bibinfo
  {journal} {Nature Reviews Materials}\ }\textbf {\bibinfo {volume} {2}},\
  \bibinfo {pages} {17031 EP } (\bibinfo {year} {2017})}\BibitemShut {NoStop}%
\bibitem [{\citenamefont {Fert}\ \emph {et~al.}(2013)\citenamefont {Fert},
  \citenamefont {Cros},\ and\ \citenamefont {Sampaio}}]{fert2013skyrmions}%
  \BibitemOpen
  \bibfield  {author} {\bibinfo {author} {\bibfnamefont {A.}~\bibnamefont
  {Fert}}, \bibinfo {author} {\bibfnamefont {V.}~\bibnamefont {Cros}}, \ and\
  \bibinfo {author} {\bibfnamefont {J.}~\bibnamefont {Sampaio}},\ }\href
  {https://doi.org/10.1038/nnano.2013.29} {\bibfield  {journal} {\bibinfo
  {journal} {Nature Nanotechnology}\ }\textbf {\bibinfo {volume} {8}},\
  \bibinfo {pages} {152 EP } (\bibinfo {year} {2013})}\BibitemShut {NoStop}%
\bibitem [{\citenamefont {Zhou}(2018)}]{zhou2018magnetic}%
  \BibitemOpen
  \bibfield  {author} {\bibinfo {author} {\bibfnamefont {Y.}~\bibnamefont
  {Zhou}},\ }\href {\doibase 10.1093/nsr/nwy109} {\bibfield  {journal}
  {\bibinfo  {journal} {National Science Review}\ }\textbf {\bibinfo {volume}
  {6}},\ \bibinfo {pages} {210} (\bibinfo {year} {2018})}\BibitemShut {NoStop}%
\bibitem [{\citenamefont {Braunecker}\ \emph {et~al.}(2010)\citenamefont
  {Braunecker}, \citenamefont {Japaridze}, \citenamefont {Klinovaja},\ and\
  \citenamefont {Loss}}]{braunecker2010spin-selective}%
  \BibitemOpen
  \bibfield  {author} {\bibinfo {author} {\bibfnamefont {B.}~\bibnamefont
  {Braunecker}}, \bibinfo {author} {\bibfnamefont {G.~I.}\ \bibnamefont
  {Japaridze}}, \bibinfo {author} {\bibfnamefont {J.}~\bibnamefont
  {Klinovaja}}, \ and\ \bibinfo {author} {\bibfnamefont {D.}~\bibnamefont
  {Loss}},\ }\href {\doibase 10.1103/PhysRevB.82.045127} {\bibfield  {journal}
  {\bibinfo  {journal} {Phys. Rev. B}\ }\textbf {\bibinfo {volume} {82}},\
  \bibinfo {pages} {045127} (\bibinfo {year} {2010})}\BibitemShut {NoStop}%
\bibitem [{\citenamefont {Klinovaja}\ and\ \citenamefont
  {Loss}(2013)}]{klinovaja2013spintronics}%
  \BibitemOpen
  \bibfield  {author} {\bibinfo {author} {\bibfnamefont {J.}~\bibnamefont
  {Klinovaja}}\ and\ \bibinfo {author} {\bibfnamefont {D.}~\bibnamefont
  {Loss}},\ }\href {\doibase 10.1103/PhysRevB.88.075404} {\bibfield  {journal}
  {\bibinfo  {journal} {Phys. Rev. B}\ }\textbf {\bibinfo {volume} {88}},\
  \bibinfo {pages} {075404} (\bibinfo {year} {2013})}\BibitemShut {NoStop}%
\bibitem [{\citenamefont {Finocchiaro}\ \emph {et~al.}(2017)\citenamefont
  {Finocchiaro}, \citenamefont {Lado},\ and\ \citenamefont
  {Fernandez-Rossier}}]{finocchiaro2017electrical}%
  \BibitemOpen
  \bibfield  {author} {\bibinfo {author} {\bibfnamefont {F.}~\bibnamefont
  {Finocchiaro}}, \bibinfo {author} {\bibfnamefont {J.~L.}\ \bibnamefont
  {Lado}}, \ and\ \bibinfo {author} {\bibfnamefont {J.}~\bibnamefont
  {Fernandez-Rossier}},\ }\href {\doibase 10.1103/PhysRevB.96.155422}
  {\bibfield  {journal} {\bibinfo  {journal} {Phys. Rev. B}\ }\textbf {\bibinfo
  {volume} {96}},\ \bibinfo {pages} {155422} (\bibinfo {year}
  {2017})}\BibitemShut {NoStop}%
\bibitem [{\citenamefont {Jonietz}\ \emph {et~al.}(2010)\citenamefont
  {Jonietz}, \citenamefont {M{\"u}hlbauer}, \citenamefont {Pfleiderer},
  \citenamefont {Neubauer}, \citenamefont {M{\"u}nzer}, \citenamefont {Bauer},
  \citenamefont {Adams}, \citenamefont {Georgii}, \citenamefont {B{\"o}ni},
  \citenamefont {Duine}, \citenamefont {Everschor}, \citenamefont {Garst},\
  and\ \citenamefont {Rosch}}]{jonietz2010spin}%
  \BibitemOpen
  \bibfield  {author} {\bibinfo {author} {\bibfnamefont {F.}~\bibnamefont
  {Jonietz}}, \bibinfo {author} {\bibfnamefont {S.}~\bibnamefont
  {M{\"u}hlbauer}}, \bibinfo {author} {\bibfnamefont {C.}~\bibnamefont
  {Pfleiderer}}, \bibinfo {author} {\bibfnamefont {A.}~\bibnamefont
  {Neubauer}}, \bibinfo {author} {\bibfnamefont {W.}~\bibnamefont
  {M{\"u}nzer}}, \bibinfo {author} {\bibfnamefont {A.}~\bibnamefont {Bauer}},
  \bibinfo {author} {\bibfnamefont {T.}~\bibnamefont {Adams}}, \bibinfo
  {author} {\bibfnamefont {R.}~\bibnamefont {Georgii}}, \bibinfo {author}
  {\bibfnamefont {P.}~\bibnamefont {B{\"o}ni}}, \bibinfo {author}
  {\bibfnamefont {R.~A.}\ \bibnamefont {Duine}}, \bibinfo {author}
  {\bibfnamefont {K.}~\bibnamefont {Everschor}}, \bibinfo {author}
  {\bibfnamefont {M.}~\bibnamefont {Garst}}, \ and\ \bibinfo {author}
  {\bibfnamefont {A.}~\bibnamefont {Rosch}},\ }\href@noop {} {\bibfield
  {journal} {\bibinfo  {journal} {Science}\ }\textbf {\bibinfo {volume}
  {330}},\ \bibinfo {pages} {1648} (\bibinfo {year} {2010})}\BibitemShut
  {NoStop}%
\bibitem [{\citenamefont {Yu}\ \emph {et~al.}(2012)\citenamefont {Yu},
  \citenamefont {Kanazawa}, \citenamefont {Zhang}, \citenamefont {Nagai},
  \citenamefont {Hara}, \citenamefont {Kimoto}, \citenamefont {Matsui},
  \citenamefont {Onose},\ and\ \citenamefont {Tokura}}]{yu2012skyrmion}%
  \BibitemOpen
  \bibfield  {author} {\bibinfo {author} {\bibfnamefont {X.~Z.}\ \bibnamefont
  {Yu}}, \bibinfo {author} {\bibfnamefont {N.}~\bibnamefont {Kanazawa}},
  \bibinfo {author} {\bibfnamefont {W.~Z.}\ \bibnamefont {Zhang}}, \bibinfo
  {author} {\bibfnamefont {T.}~\bibnamefont {Nagai}}, \bibinfo {author}
  {\bibfnamefont {T.}~\bibnamefont {Hara}}, \bibinfo {author} {\bibfnamefont
  {K.}~\bibnamefont {Kimoto}}, \bibinfo {author} {\bibfnamefont
  {Y.}~\bibnamefont {Matsui}}, \bibinfo {author} {\bibfnamefont
  {Y.}~\bibnamefont {Onose}}, \ and\ \bibinfo {author} {\bibfnamefont
  {Y.}~\bibnamefont {Tokura}},\ }\href {\doibase 10.1038/ncomms1990} {\bibfield
   {journal} {\bibinfo  {journal} {Nature Communications}\ }\textbf {\bibinfo
  {volume} {3}},\ \bibinfo {pages} {988} (\bibinfo {year} {2012})}\BibitemShut
  {NoStop}%
\bibitem [{\citenamefont {Romming}\ \emph {et~al.}(2013)\citenamefont
  {Romming}, \citenamefont {Hanneken}, \citenamefont {Menzel}, \citenamefont
  {Bickel}, \citenamefont {Wolter}, \citenamefont {von Bergmann}, \citenamefont
  {Kubetzka},\ and\ \citenamefont {Wiesendanger}}]{romming2013writing}%
  \BibitemOpen
  \bibfield  {author} {\bibinfo {author} {\bibfnamefont {N.}~\bibnamefont
  {Romming}}, \bibinfo {author} {\bibfnamefont {C.}~\bibnamefont {Hanneken}},
  \bibinfo {author} {\bibfnamefont {M.}~\bibnamefont {Menzel}}, \bibinfo
  {author} {\bibfnamefont {J.~E.}\ \bibnamefont {Bickel}}, \bibinfo {author}
  {\bibfnamefont {B.}~\bibnamefont {Wolter}}, \bibinfo {author} {\bibfnamefont
  {K.}~\bibnamefont {von Bergmann}}, \bibinfo {author} {\bibfnamefont
  {A.}~\bibnamefont {Kubetzka}}, \ and\ \bibinfo {author} {\bibfnamefont
  {R.}~\bibnamefont {Wiesendanger}},\ }\href@noop {} {\bibfield  {journal}
  {\bibinfo  {journal} {Science}\ }\textbf {\bibinfo {volume} {341}},\ \bibinfo
  {pages} {636} (\bibinfo {year} {2013})}\BibitemShut {NoStop}%
\bibitem [{\citenamefont {Moreau-Luchaire}\ \emph {et~al.}(2016)\citenamefont
  {Moreau-Luchaire}, \citenamefont {Moutafis}, \citenamefont {Reyren},
  \citenamefont {Sampaio}, \citenamefont {Vaz}, \citenamefont {Van~Horne},
  \citenamefont {Bouzehouane}, \citenamefont {Garcia}, \citenamefont
  {Deranlot}, \citenamefont {Warnicke}, \citenamefont {Wohlh{\"u}ter},
  \citenamefont {George}, \citenamefont {Weigand}, \citenamefont {Raabe},
  \citenamefont {Cros},\ and\ \citenamefont
  {Fert}}]{moreau-uchaire2016additive}%
  \BibitemOpen
  \bibfield  {author} {\bibinfo {author} {\bibfnamefont {C.}~\bibnamefont
  {Moreau-Luchaire}}, \bibinfo {author} {\bibfnamefont {C.}~\bibnamefont
  {Moutafis}}, \bibinfo {author} {\bibfnamefont {N.}~\bibnamefont {Reyren}},
  \bibinfo {author} {\bibfnamefont {J.}~\bibnamefont {Sampaio}}, \bibinfo
  {author} {\bibfnamefont {C.~A.~F.}\ \bibnamefont {Vaz}}, \bibinfo {author}
  {\bibfnamefont {N.}~\bibnamefont {Van~Horne}}, \bibinfo {author}
  {\bibfnamefont {K.}~\bibnamefont {Bouzehouane}}, \bibinfo {author}
  {\bibfnamefont {K.}~\bibnamefont {Garcia}}, \bibinfo {author} {\bibfnamefont
  {C.}~\bibnamefont {Deranlot}}, \bibinfo {author} {\bibfnamefont
  {P.}~\bibnamefont {Warnicke}}, \bibinfo {author} {\bibfnamefont
  {P.}~\bibnamefont {Wohlh{\"u}ter}}, \bibinfo {author} {\bibfnamefont {J.~M.}\
  \bibnamefont {George}}, \bibinfo {author} {\bibfnamefont {M.}~\bibnamefont
  {Weigand}}, \bibinfo {author} {\bibfnamefont {J.}~\bibnamefont {Raabe}},
  \bibinfo {author} {\bibfnamefont {V.}~\bibnamefont {Cros}}, \ and\ \bibinfo
  {author} {\bibfnamefont {A.}~\bibnamefont {Fert}},\ }\href
  {https://doi.org/10.1038/nnano.2015.313} {\bibfield  {journal} {\bibinfo
  {journal} {Nature Nanotechnology}\ }\textbf {\bibinfo {volume} {11}},\
  \bibinfo {pages} {444 EP } (\bibinfo {year} {2016})}\BibitemShut {NoStop}%
\bibitem [{\citenamefont {Ajejas}\ \emph {et~al.}(2018)\citenamefont {Ajejas},
  \citenamefont {Gud{\'\i}n}, \citenamefont {Guerrero}, \citenamefont
  {Anad{\'o}n~Barcelona}, \citenamefont {Diez}, \citenamefont {de~Melo~Costa},
  \citenamefont {Olleros}, \citenamefont {Ni{\~n}o}, \citenamefont {Pizzini},
  \citenamefont {Vogel}, \citenamefont {Valvidares}, \citenamefont {Gargiani},
  \citenamefont {Cabero}, \citenamefont {Varela}, \citenamefont {Camarero},
  \citenamefont {Miranda},\ and\ \citenamefont {Perna}}]{ajejas2018unraveling}%
  \BibitemOpen
  \bibfield  {author} {\bibinfo {author} {\bibfnamefont {F.}~\bibnamefont
  {Ajejas}}, \bibinfo {author} {\bibfnamefont {A.}~\bibnamefont {Gud{\'\i}n}},
  \bibinfo {author} {\bibfnamefont {R.}~\bibnamefont {Guerrero}}, \bibinfo
  {author} {\bibfnamefont {A.}~\bibnamefont {Anad{\'o}n~Barcelona}}, \bibinfo
  {author} {\bibfnamefont {J.~M.}\ \bibnamefont {Diez}}, \bibinfo {author}
  {\bibfnamefont {L.}~\bibnamefont {de~Melo~Costa}}, \bibinfo {author}
  {\bibfnamefont {P.}~\bibnamefont {Olleros}}, \bibinfo {author} {\bibfnamefont
  {M.~A.}\ \bibnamefont {Ni{\~n}o}}, \bibinfo {author} {\bibfnamefont
  {S.}~\bibnamefont {Pizzini}}, \bibinfo {author} {\bibfnamefont
  {J.}~\bibnamefont {Vogel}}, \bibinfo {author} {\bibfnamefont
  {M.}~\bibnamefont {Valvidares}}, \bibinfo {author} {\bibfnamefont
  {P.}~\bibnamefont {Gargiani}}, \bibinfo {author} {\bibfnamefont
  {M.}~\bibnamefont {Cabero}}, \bibinfo {author} {\bibfnamefont
  {M.}~\bibnamefont {Varela}}, \bibinfo {author} {\bibfnamefont
  {J.}~\bibnamefont {Camarero}}, \bibinfo {author} {\bibfnamefont
  {R.}~\bibnamefont {Miranda}}, \ and\ \bibinfo {author} {\bibfnamefont
  {P.}~\bibnamefont {Perna}},\ }\href {\doibase 10.1021/acs.nanolett.8b00878}
  {\bibfield  {journal} {\bibinfo  {journal} {Nano Letters}\ }\textbf {\bibinfo
  {volume} {18}},\ \bibinfo {pages} {5364} (\bibinfo {year}
  {2018})}\BibitemShut {NoStop}%
\bibitem [{\citenamefont {Ajejas}\ \emph {et~al.}(2017)\citenamefont {Ajejas},
  \citenamefont {K{\v r}i{\v z}{\'a}kov{\'a}}, \citenamefont {de~Souza~Chaves},
  \citenamefont {Vogel}, \citenamefont {Perna}, \citenamefont {Guerrero},
  \citenamefont {Gudin}, \citenamefont {Camarero},\ and\ \citenamefont
  {Pizzini}}]{ajejas2017tuning}%
  \BibitemOpen
  \bibfield  {author} {\bibinfo {author} {\bibfnamefont {F.}~\bibnamefont
  {Ajejas}}, \bibinfo {author} {\bibfnamefont {V.}~\bibnamefont {K{\v r}i{\v
  z}{\'a}kov{\'a}}}, \bibinfo {author} {\bibfnamefont {D.}~\bibnamefont
  {de~Souza~Chaves}}, \bibinfo {author} {\bibfnamefont {J.}~\bibnamefont
  {Vogel}}, \bibinfo {author} {\bibfnamefont {P.}~\bibnamefont {Perna}},
  \bibinfo {author} {\bibfnamefont {R.}~\bibnamefont {Guerrero}}, \bibinfo
  {author} {\bibfnamefont {A.}~\bibnamefont {Gudin}}, \bibinfo {author}
  {\bibfnamefont {J.}~\bibnamefont {Camarero}}, \ and\ \bibinfo {author}
  {\bibfnamefont {S.}~\bibnamefont {Pizzini}},\ }\href {\doibase
  10.1063/1.5005798} {\bibfield  {journal} {\bibinfo  {journal} {Applied
  Physics Letters}\ }\textbf {\bibinfo {volume} {111}},\ \bibinfo {pages}
  {202402} (\bibinfo {year} {2017})}\BibitemShut {NoStop}%
\bibitem [{\citenamefont {Pellegrino}\ \emph {et~al.}(2014)\citenamefont
  {Pellegrino}, \citenamefont {Chirolli}, \citenamefont {Fazio}, \citenamefont
  {Giovannetti},\ and\ \citenamefont {Polini}}]{pellegrino2014theory}%
  \BibitemOpen
  \bibfield  {author} {\bibinfo {author} {\bibfnamefont {F.~M.~D.}\
  \bibnamefont {Pellegrino}}, \bibinfo {author} {\bibfnamefont
  {L.}~\bibnamefont {Chirolli}}, \bibinfo {author} {\bibfnamefont
  {R.}~\bibnamefont {Fazio}}, \bibinfo {author} {\bibfnamefont
  {V.}~\bibnamefont {Giovannetti}}, \ and\ \bibinfo {author} {\bibfnamefont
  {M.}~\bibnamefont {Polini}},\ }\href {\doibase 10.1103/PhysRevB.89.165406}
  {\bibfield  {journal} {\bibinfo  {journal} {Phys. Rev. B}\ }\textbf {\bibinfo
  {volume} {89}},\ \bibinfo {pages} {165406} (\bibinfo {year}
  {2014})}\BibitemShut {NoStop}%
\bibitem [{\citenamefont {Glazov}\ \emph {et~al.}(2017)\citenamefont {Glazov},
  \citenamefont {Golub}, \citenamefont {Wang}, \citenamefont {Marie},
  \citenamefont {Amand},\ and\ \citenamefont {Urbaszek}}]{glazov2017intrinsic}%
  \BibitemOpen
  \bibfield  {author} {\bibinfo {author} {\bibfnamefont {M.~M.}\ \bibnamefont
  {Glazov}}, \bibinfo {author} {\bibfnamefont {L.~E.}\ \bibnamefont {Golub}},
  \bibinfo {author} {\bibfnamefont {G.}~\bibnamefont {Wang}}, \bibinfo {author}
  {\bibfnamefont {X.}~\bibnamefont {Marie}}, \bibinfo {author} {\bibfnamefont
  {T.}~\bibnamefont {Amand}}, \ and\ \bibinfo {author} {\bibfnamefont
  {B.}~\bibnamefont {Urbaszek}},\ }\href {\doibase 10.1103/PhysRevB.95.035311}
  {\bibfield  {journal} {\bibinfo  {journal} {Phys. Rev. B}\ }\textbf {\bibinfo
  {volume} {95}},\ \bibinfo {pages} {035311} (\bibinfo {year}
  {2017})}\BibitemShut {NoStop}%
\bibitem [{\citenamefont {Korm\'anyos}\ \emph {et~al.}(2013)\citenamefont
  {Korm\'anyos}, \citenamefont {Z\'olyomi}, \citenamefont {Drummond},
  \citenamefont {Rakyta}, \citenamefont {Burkard},\ and\ \citenamefont
  {Fal'ko}}]{kormanyos2013monolayer}%
  \BibitemOpen
  \bibfield  {author} {\bibinfo {author} {\bibfnamefont {A.}~\bibnamefont
  {Korm\'anyos}}, \bibinfo {author} {\bibfnamefont {V.}~\bibnamefont
  {Z\'olyomi}}, \bibinfo {author} {\bibfnamefont {N.~D.}\ \bibnamefont
  {Drummond}}, \bibinfo {author} {\bibfnamefont {P.}~\bibnamefont {Rakyta}},
  \bibinfo {author} {\bibfnamefont {G.}~\bibnamefont {Burkard}}, \ and\
  \bibinfo {author} {\bibfnamefont {V.~I.}\ \bibnamefont {Fal'ko}},\ }\href
  {\doibase 10.1103/PhysRevB.88.045416} {\bibfield  {journal} {\bibinfo
  {journal} {Phys. Rev. B}\ }\textbf {\bibinfo {volume} {88}},\ \bibinfo
  {pages} {045416} (\bibinfo {year} {2013})}\BibitemShut {NoStop}%
\bibitem [{\citenamefont {Korm{\'a}nyos}\ \emph {et~al.}(2015)\citenamefont
  {Korm{\'a}nyos}, \citenamefont {Burkard}, \citenamefont {Gmitra},
  \citenamefont {Fabian}, \citenamefont {Z{\'o}lyomi}, \citenamefont
  {Drummond},\ and\ \citenamefont {Fal'ko}}]{kormanyos2015kp}%
  \BibitemOpen
  \bibfield  {author} {\bibinfo {author} {\bibfnamefont {A.}~\bibnamefont
  {Korm{\'a}nyos}}, \bibinfo {author} {\bibfnamefont {G.}~\bibnamefont
  {Burkard}}, \bibinfo {author} {\bibfnamefont {M.}~\bibnamefont {Gmitra}},
  \bibinfo {author} {\bibfnamefont {J.}~\bibnamefont {Fabian}}, \bibinfo
  {author} {\bibfnamefont {V.}~\bibnamefont {Z{\'o}lyomi}}, \bibinfo {author}
  {\bibfnamefont {N.~D.}\ \bibnamefont {Drummond}}, \ and\ \bibinfo {author}
  {\bibfnamefont {V.}~\bibnamefont {Fal'ko}},\ }\href
  {http://stacks.iop.org/2053-1583/2/i=2/a=022001} {\bibfield  {journal}
  {\bibinfo  {journal} {2D Materials}\ }\textbf {\bibinfo {volume} {2}},\
  \bibinfo {pages} {022001} (\bibinfo {year} {2015})}\BibitemShut {NoStop}%
\bibitem [{\citenamefont {Yang}\ \emph {et~al.}(1991)\citenamefont {Yang},
  \citenamefont {Guo}, \citenamefont {Chan}, \citenamefont {Wong},\ and\
  \citenamefont {Ching}}]{yang1991analytic}%
  \BibitemOpen
  \bibfield  {author} {\bibinfo {author} {\bibfnamefont {X.~L.}\ \bibnamefont
  {Yang}}, \bibinfo {author} {\bibfnamefont {S.~H.}\ \bibnamefont {Guo}},
  \bibinfo {author} {\bibfnamefont {F.~T.}\ \bibnamefont {Chan}}, \bibinfo
  {author} {\bibfnamefont {K.~W.}\ \bibnamefont {Wong}}, \ and\ \bibinfo
  {author} {\bibfnamefont {W.~Y.}\ \bibnamefont {Ching}},\ }\href {\doibase
  10.1103/PhysRevA.43.1186} {\bibfield  {journal} {\bibinfo  {journal}
  {Physical Review A}\ }\textbf {\bibinfo {volume} {43}},\ \bibinfo {pages}
  {1186} (\bibinfo {year} {1991})}\BibitemShut {NoStop}%
\bibitem [{\citenamefont {Prada}\ \emph {et~al.}(2015)\citenamefont {Prada},
  \citenamefont {Alvarez}, \citenamefont {Narasimha-Acharya}, \citenamefont
  {Bailen},\ and\ \citenamefont {Palacios}}]{prada2015effective-mass}%
  \BibitemOpen
  \bibfield  {author} {\bibinfo {author} {\bibfnamefont {E.}~\bibnamefont
  {Prada}}, \bibinfo {author} {\bibfnamefont {J.~V.}\ \bibnamefont {Alvarez}},
  \bibinfo {author} {\bibfnamefont {K.~L.}\ \bibnamefont {Narasimha-Acharya}},
  \bibinfo {author} {\bibfnamefont {F.~J.}\ \bibnamefont {Bailen}}, \ and\
  \bibinfo {author} {\bibfnamefont {J.~J.}\ \bibnamefont {Palacios}},\ }\href
  {\doibase 10.1103/PhysRevB.91.245421} {\bibfield  {journal} {\bibinfo
  {journal} {Physical Review B}\ }\textbf {\bibinfo {volume} {91}},\ \bibinfo
  {pages} {245421} (\bibinfo {year} {2015})}\BibitemShut {NoStop}%
\bibitem [{\citenamefont {Echeverry}\ \emph {et~al.}(2016)\citenamefont
  {Echeverry}, \citenamefont {Urbaszek}, \citenamefont {Amand}, \citenamefont
  {Marie},\ and\ \citenamefont {Gerber}}]{echeverry2016splitting}%
  \BibitemOpen
  \bibfield  {author} {\bibinfo {author} {\bibfnamefont {J.~P.}\ \bibnamefont
  {Echeverry}}, \bibinfo {author} {\bibfnamefont {B.}~\bibnamefont {Urbaszek}},
  \bibinfo {author} {\bibfnamefont {T.}~\bibnamefont {Amand}}, \bibinfo
  {author} {\bibfnamefont {X.}~\bibnamefont {Marie}}, \ and\ \bibinfo {author}
  {\bibfnamefont {I.~C.}\ \bibnamefont {Gerber}},\ }\href {\doibase
  10.1103/PhysRevB.93.121107} {\bibfield  {journal} {\bibinfo  {journal} {Phys.
  Rev. B}\ }\textbf {\bibinfo {volume} {93}},\ \bibinfo {pages} {121107}
  (\bibinfo {year} {2016})}\BibitemShut {NoStop}%
\bibitem [{\citenamefont {Yu}\ \emph {et~al.}(2014)\citenamefont {Yu},
  \citenamefont {Liu}, \citenamefont {Gong}, \citenamefont {Xu},\ and\
  \citenamefont {Yao}}]{yu2014dirac}%
  \BibitemOpen
  \bibfield  {author} {\bibinfo {author} {\bibfnamefont {H.}~\bibnamefont
  {Yu}}, \bibinfo {author} {\bibfnamefont {G.-B.}\ \bibnamefont {Liu}},
  \bibinfo {author} {\bibfnamefont {P.}~\bibnamefont {Gong}}, \bibinfo {author}
  {\bibfnamefont {X.}~\bibnamefont {Xu}}, \ and\ \bibinfo {author}
  {\bibfnamefont {W.}~\bibnamefont {Yao}},\ }\href {\doibase
  10.1038/ncomms4876} {\bibfield  {journal} {\bibinfo  {journal} {Nature
  Communications}\ }\textbf {\bibinfo {volume} {5}},\ \bibinfo {pages} {1}
  (\bibinfo {year} {2014})},\ \Eprint {http://arxiv.org/abs/1401.0667}
  {1401.0667} \BibitemShut {NoStop}%
\bibitem [{\citenamefont {Slobodeniuk}\ and\ \citenamefont
  {Basko}(2016)}]{slobodeniuk2016spin-flip}%
  \BibitemOpen
  \bibfield  {author} {\bibinfo {author} {\bibfnamefont {A.~O.}\ \bibnamefont
  {Slobodeniuk}}\ and\ \bibinfo {author} {\bibfnamefont {D.~M.}\ \bibnamefont
  {Basko}},\ }\href {\doibase 10.1088/2053-1583/3/3/035009} {\bibfield
  {journal} {\bibinfo  {journal} {2D Materials}\ }\textbf {\bibinfo {volume}
  {3}},\ \bibinfo {pages} {035009} (\bibinfo {year} {2016})}\BibitemShut
  {NoStop}%
\bibitem [{\citenamefont {Karmakar}\ \emph {et~al.}(2011)\citenamefont
  {Karmakar}, \citenamefont {Venturelli}, \citenamefont {Chirolli},
  \citenamefont {Taddei}, \citenamefont {Giovannetti}, \citenamefont {Fazio},
  \citenamefont {Roddaro}, \citenamefont {Biasiol}, \citenamefont {Sorba},
  \citenamefont {Pellegrini},\ and\ \citenamefont
  {Beltram}}]{karmakar2011controlled}%
  \BibitemOpen
  \bibfield  {author} {\bibinfo {author} {\bibfnamefont {B.}~\bibnamefont
  {Karmakar}}, \bibinfo {author} {\bibfnamefont {D.}~\bibnamefont
  {Venturelli}}, \bibinfo {author} {\bibfnamefont {L.}~\bibnamefont
  {Chirolli}}, \bibinfo {author} {\bibfnamefont {F.}~\bibnamefont {Taddei}},
  \bibinfo {author} {\bibfnamefont {V.}~\bibnamefont {Giovannetti}}, \bibinfo
  {author} {\bibfnamefont {R.}~\bibnamefont {Fazio}}, \bibinfo {author}
  {\bibfnamefont {S.}~\bibnamefont {Roddaro}}, \bibinfo {author} {\bibfnamefont
  {G.}~\bibnamefont {Biasiol}}, \bibinfo {author} {\bibfnamefont
  {L.}~\bibnamefont {Sorba}}, \bibinfo {author} {\bibfnamefont
  {V.}~\bibnamefont {Pellegrini}}, \ and\ \bibinfo {author} {\bibfnamefont
  {F.}~\bibnamefont {Beltram}},\ }\href {\doibase
  10.1103/PhysRevLett.107.236804} {\bibfield  {journal} {\bibinfo  {journal}
  {Phys. Rev. Lett.}\ }\textbf {\bibinfo {volume} {107}},\ \bibinfo {pages}
  {236804} (\bibinfo {year} {2011})}\BibitemShut {NoStop}%
\bibitem [{\citenamefont {Kjaergaard}\ \emph {et~al.}(2012)\citenamefont
  {Kjaergaard}, \citenamefont {W\"olms},\ and\ \citenamefont
  {Flensberg}}]{kjaergaard2012majorana}%
  \BibitemOpen
  \bibfield  {author} {\bibinfo {author} {\bibfnamefont {M.}~\bibnamefont
  {Kjaergaard}}, \bibinfo {author} {\bibfnamefont {K.}~\bibnamefont {W\"olms}},
  \ and\ \bibinfo {author} {\bibfnamefont {K.}~\bibnamefont {Flensberg}},\
  }\href {\doibase 10.1103/PhysRevB.85.020503} {\bibfield  {journal} {\bibinfo
  {journal} {Phys. Rev. B}\ }\textbf {\bibinfo {volume} {85}},\ \bibinfo
  {pages} {020503} (\bibinfo {year} {2012})}\BibitemShut {NoStop}%
\bibitem [{\citenamefont {Robert}\ \emph {et~al.}(2017)\citenamefont {Robert},
  \citenamefont {Amand}, \citenamefont {Cadiz}, \citenamefont {Lagarde},
  \citenamefont {Courtade}, \citenamefont {Manca}, \citenamefont {Taniguchi},
  \citenamefont {Watanabe}, \citenamefont {Urbaszek},\ and\ \citenamefont
  {Marie}}]{robert2017fine}%
  \BibitemOpen
  \bibfield  {author} {\bibinfo {author} {\bibfnamefont {C.}~\bibnamefont
  {Robert}}, \bibinfo {author} {\bibfnamefont {T.}~\bibnamefont {Amand}},
  \bibinfo {author} {\bibfnamefont {F.}~\bibnamefont {Cadiz}}, \bibinfo
  {author} {\bibfnamefont {D.}~\bibnamefont {Lagarde}}, \bibinfo {author}
  {\bibfnamefont {E.}~\bibnamefont {Courtade}}, \bibinfo {author}
  {\bibfnamefont {M.}~\bibnamefont {Manca}}, \bibinfo {author} {\bibfnamefont
  {T.}~\bibnamefont {Taniguchi}}, \bibinfo {author} {\bibfnamefont
  {K.}~\bibnamefont {Watanabe}}, \bibinfo {author} {\bibfnamefont
  {B.}~\bibnamefont {Urbaszek}}, \ and\ \bibinfo {author} {\bibfnamefont
  {X.}~\bibnamefont {Marie}},\ }\href {\doibase 10.1103/PhysRevB.96.155423}
  {\bibfield  {journal} {\bibinfo  {journal} {Phys. Rev. B}\ }\textbf {\bibinfo
  {volume} {96}},\ \bibinfo {pages} {155423} (\bibinfo {year}
  {2017})}\BibitemShut {NoStop}%
\bibitem [{\citenamefont {Wang}\ \emph {et~al.}(2017)\citenamefont {Wang},
  \citenamefont {Robert}, \citenamefont {Glazov}, \citenamefont {Cadiz},
  \citenamefont {Courtade}, \citenamefont {Amand}, \citenamefont {Lagarde},
  \citenamefont {Taniguchi}, \citenamefont {Watanabe}, \citenamefont
  {Urbaszek},\ and\ \citenamefont {Marie}}]{wang2017in-plane}%
  \BibitemOpen
  \bibfield  {author} {\bibinfo {author} {\bibfnamefont {G.}~\bibnamefont
  {Wang}}, \bibinfo {author} {\bibfnamefont {C.}~\bibnamefont {Robert}},
  \bibinfo {author} {\bibfnamefont {M.~M.}\ \bibnamefont {Glazov}}, \bibinfo
  {author} {\bibfnamefont {F.}~\bibnamefont {Cadiz}}, \bibinfo {author}
  {\bibfnamefont {E.}~\bibnamefont {Courtade}}, \bibinfo {author}
  {\bibfnamefont {T.}~\bibnamefont {Amand}}, \bibinfo {author} {\bibfnamefont
  {D.}~\bibnamefont {Lagarde}}, \bibinfo {author} {\bibfnamefont
  {T.}~\bibnamefont {Taniguchi}}, \bibinfo {author} {\bibfnamefont
  {K.}~\bibnamefont {Watanabe}}, \bibinfo {author} {\bibfnamefont
  {B.}~\bibnamefont {Urbaszek}}, \ and\ \bibinfo {author} {\bibfnamefont
  {X.}~\bibnamefont {Marie}},\ }\href {\doibase 10.1103/PhysRevLett.119.047401}
  {\bibfield  {journal} {\bibinfo  {journal} {Phys. Rev. Lett.}\ }\textbf
  {\bibinfo {volume} {119}},\ \bibinfo {pages} {047401} (\bibinfo {year}
  {2017})}\BibitemShut {NoStop}%
\bibitem [{\citenamefont {Molas}\ \emph {et~al.}(2017)\citenamefont {Molas},
  \citenamefont {Faugeras}, \citenamefont {Slobodeniuk}, \citenamefont
  {Nogajewski}, \citenamefont {Bartos}, \citenamefont {Basko},\ and\
  \citenamefont {Potemski}}]{molas2017brightening}%
  \BibitemOpen
  \bibfield  {author} {\bibinfo {author} {\bibfnamefont {M.~R.}\ \bibnamefont
  {Molas}}, \bibinfo {author} {\bibfnamefont {C.}~\bibnamefont {Faugeras}},
  \bibinfo {author} {\bibfnamefont {A.~O.}\ \bibnamefont {Slobodeniuk}},
  \bibinfo {author} {\bibfnamefont {K.}~\bibnamefont {Nogajewski}}, \bibinfo
  {author} {\bibfnamefont {M.}~\bibnamefont {Bartos}}, \bibinfo {author}
  {\bibfnamefont {D.~M.}\ \bibnamefont {Basko}}, \ and\ \bibinfo {author}
  {\bibfnamefont {M.}~\bibnamefont {Potemski}},\ }\href {\doibase
  10.1088/2053-1583/aa5521} {\bibfield  {journal} {\bibinfo  {journal} {2D
  Materials}\ }\textbf {\bibinfo {volume} {4}},\ \bibinfo {pages} {021003}
  (\bibinfo {year} {2017})}\BibitemShut {NoStop}%
\bibitem [{\citenamefont {Dery}\ and\ \citenamefont
  {Song}(2015)}]{dery2015polarization}%
  \BibitemOpen
  \bibfield  {author} {\bibinfo {author} {\bibfnamefont {H.}~\bibnamefont
  {Dery}}\ and\ \bibinfo {author} {\bibfnamefont {Y.}~\bibnamefont {Song}},\
  }\href {\doibase 10.1103/PhysRevB.92.125431} {\bibfield  {journal} {\bibinfo
  {journal} {Phys. Rev. B}\ }\textbf {\bibinfo {volume} {92}},\ \bibinfo
  {pages} {125431} (\bibinfo {year} {2015})}\BibitemShut {NoStop}%
\bibitem [{\citenamefont {Molas}\ \emph {et~al.}(2019)\citenamefont {Molas},
  \citenamefont {Slobodeniuk}, \citenamefont {Kazimierczuk}, \citenamefont
  {Nogajewski}, \citenamefont {Bartos}, \citenamefont {Kapu\ifmmode
  \acute{s}\else \'{s}\fi{}ci\ifmmode~\acute{n}\else \'{n}\fi{}ski},
  \citenamefont {Oreszczuk}, \citenamefont {Watanabe}, \citenamefont
  {Taniguchi}, \citenamefont {Faugeras}, \citenamefont {Kossacki},
  \citenamefont {Basko},\ and\ \citenamefont {Potemski}}]{molas2019probing}%
  \BibitemOpen
  \bibfield  {author} {\bibinfo {author} {\bibfnamefont {M.~R.}\ \bibnamefont
  {Molas}}, \bibinfo {author} {\bibfnamefont {A.~O.}\ \bibnamefont
  {Slobodeniuk}}, \bibinfo {author} {\bibfnamefont {T.}~\bibnamefont
  {Kazimierczuk}}, \bibinfo {author} {\bibfnamefont {K.}~\bibnamefont
  {Nogajewski}}, \bibinfo {author} {\bibfnamefont {M.}~\bibnamefont {Bartos}},
  \bibinfo {author} {\bibfnamefont {P.}~\bibnamefont {Kapu\ifmmode
  \acute{s}\else \'{s}\fi{}ci\ifmmode~\acute{n}\else \'{n}\fi{}ski}}, \bibinfo
  {author} {\bibfnamefont {K.}~\bibnamefont {Oreszczuk}}, \bibinfo {author}
  {\bibfnamefont {K.}~\bibnamefont {Watanabe}}, \bibinfo {author}
  {\bibfnamefont {T.}~\bibnamefont {Taniguchi}}, \bibinfo {author}
  {\bibfnamefont {C.}~\bibnamefont {Faugeras}}, \bibinfo {author}
  {\bibfnamefont {P.}~\bibnamefont {Kossacki}}, \bibinfo {author}
  {\bibfnamefont {D.~M.}\ \bibnamefont {Basko}}, \ and\ \bibinfo {author}
  {\bibfnamefont {M.}~\bibnamefont {Potemski}},\ }\href {\doibase
  10.1103/PhysRevLett.123.096803} {\bibfield  {journal} {\bibinfo  {journal}
  {Phys. Rev. Lett.}\ }\textbf {\bibinfo {volume} {123}},\ \bibinfo {pages}
  {096803} (\bibinfo {year} {2019})}\BibitemShut {NoStop}%
\end{thebibliography}%

\end{document}